\documentstyle[epsfig,natbib2,natbibmnfix]{mn}

\newcommand{\be}{\begin{equation}}
\newcommand{\ee}{\end{equation}}
\newcommand{\bea}{\begin{eqnarray}}
\newcommand{\eea}{\end{eqnarray}}
\newcommand{\bc}{\begin{center}}
\newcommand{\ec}{\end{center}}

\def\ltsima{$\; \buildrel < \over \sim \;$}
\def\simlt{\lower.5ex\hbox{\ltsima}}
\def\gtsima{$\; \buildrel > \over \sim \;$}
\def\simgt{\lower.5ex\hbox{\gtsima}}

\renewcommand{\thefootnote}{\fnsymbol{footnote}}

\title{An analytical model for the history of cosmic star formation}

\author[L.~Hernquist and V.~Springel] {\parbox{18cm}{Lars
Hernquist$^1$\footnotemark[1] and Volker
Springel$^{2}$\footnotemark[3]}\vspace{0.3cm}\\
$^1$Harvard-Smithsonian Center for Astrophysics, 60 Garden Street,
Cambridge, MA 02138, USA\\ $^2$Max-Planck-Institut f\"{u}r
Astrophysik, Karl-Schwarzschild-Stra\ss{}e 1, 85740 Garching bei
M\"{u}nchen, Germany}

\begin{document}

\maketitle
\begin{abstract}

We use simple analytic reasoning to identify physical processes that
drive the evolution of the cosmic star formation rate, $\dot\rho_\star$,
in cold dark matter universes.  Based on our analysis, we formulate a 
model to characterise the redshift dependence of $\dot\rho_\star$ and
compare it to results obtained from a set of hydrodynamic simulations
which include star formation and feedback.

We find that the cosmic star formation rate is described by two
regimes.  At early times, densities are sufficiently high and cooling
times sufficiently short that abundant quantities of star-forming gas
are present in all dark matter halos that can cool by atomic
processes.  Consequently, $\dot\rho_\star$ generically rises
exponentially as $z$ decreases, independent of the details of the
physical model for star formation, but dependent on the normalisation
and shape of the cosmological power spectrum.  This part of the
evolution is dominated by gravitationally driven growth of the halo
mass function.  At low redshifts, densities decline as the universe
expands to the point that cooling is inhibited, limiting the amount of
star-forming gas available.  We find that in this regime the star
formation rate scales approximately as $\dot\rho_\star\propto
H(z)^{4/3}$, in proportion to the cooling rate within halos.

We demonstrate that the existence of these two regimes leads
to a peak in the star formation rate at an intermediate redshift $z =
z_{\rm peak}$.  We discuss how the location of this peak depends on
our model parameters.  Only star formation efficiencies that are
unrealistically low would delay the peak to $z\simeq 3$ or below, and
we show that the peak cannot occur above a limiting redshift of $z
\approx 8.7$.  For the star formation efficiency adopted in our
numerical simulations, $z_{\rm peak} \approx 5 - 6$.

We derive analytic expressions for the full star formation history and
show that they match our simulation results to better than
$\simeq$10\%.  Using various approximations, we reduce the expressions
to a simple analytic fitting function for $\dot\rho_\star$ that can be
used to compute global cosmological quantities that are directly
related to the star formation history.  As examples, we consider the
integrated stellar density, the supernova and gamma-ray burst (GRB)
rates observable on Earth, the metal enrichment history of the
Universe, and the density of compact objects.  We also briefly discuss
the expected dependence of the star formation history on cosmological
parameters and the physics of the gas.

\end{abstract}

\section{Introduction}

\renewcommand{\thefootnote}{\fnsymbol{footnote}}
\footnotetext[1]{E-mail: lars@cfa.harvard.edu}
\footnotetext[3]{\hspace{0.03cm}E-mail: volker@mpa-garching.mpg.de}

The history of cosmic star formation is of fundamental importance to
cosmology, not only to galaxy formation itself, but also for ongoing
efforts to determine cosmological parameters and the matter content of
the Universe.  Over the past decade, various attempts have been made
to directly map out the evolution of star formation observationally
\citep[e.g.][]{Gal95,Mad96,Mad98,Lil96,Cow96,Cow99,Con97,Hug98,Trey98,
Tresse98,Pas98,Steid99,Flo99,Gron99,Hogg01,Bald02,Lan2002,Wil02}.
Obtaining precise measurements of the star formation rate density,
$\dot\rho_\star$, is made challenging, however, by the difficult
nature of these observations and also by uncertainties in systematic
effects such as dust extinction.  Partly for this reason, there is
strong motivation for predicting $\dot\rho_\star$ theoretically to
provide a framework for interpreting the data.

Various theoretical efforts have been made to calculate
$\dot\rho_\star$, using either semi-analytic models
\citep{Wh91,Cole1994,Bau98,Som00}, or numerical simulations
\citep{Weinberg99,Pea2000,Nag00,Nag01,Asc02}.  Unfortunately, due to
the complexity of the physics underlying galaxy formation, the
predicted behaviour for $\dot\rho_\star(z)$ can be quite sensitive to
the model adopted to describe star formation and associated
feedback processes.  Perhaps because of this difficulty, there have
been few attempts to determine whether some aspects of the expected
evolution of the star formation density in cold dark matter
cosmologies are relatively insensitive to the details of the physics
of star formation. If such a ``generic'' behaviour exists within a
reasonably broad class of physical models, it should be possible to
make robust predictions for the shape of the star formation history in
cold dark matter universes that could be confronted with observations
to test the currently favoured paradigm of hierarchical galaxy
formation.

In this paper, we examine this issue in detail.  We are motivated by
the numerical results presented in \citet{SprHerSFR}, where we used a
large set of hydrodynamic simulations to infer the evolution of the
cosmic star formation rate density from high redshift to the present.
These simulations included a novel description for star formation and
feedback processes within the interstellar medium
\citep{SprHerMultiPhase} and a novel formulation of the equations of
motion \citep{SprHerEnt}.  The broad range of scales encompassed by
our set of simulations, together with extensive convergence tests,
enabled us to obtain a converged prediction for $\dot\rho_\star (z)$
within this model for galaxy formation.

The cosmic star formation history we inferred peaks at a redshift
$z_{\rm peak} \sim 5.5$, declining roughly exponentially towards both
low and high redshift.  Here, we establish a physical basis for the
particular form of the star formation history predicted by our
simulations.  This makes it possible to arrive at a clearer
understanding of the physics that drives the evolution of the cosmic
star formation history, and allows us to justify specific analytic
fitting functions for the full star formation history.  Such
closed-form descriptions are particularly useful for computing derived
quantities that directly depend on the star formation history and for
relating theoretical predictions to observations.

This paper is organised as follows. In Section~\ref{secfit}, we
present our analytic fitting function for the cosmic star
formation history, followed in Section~\ref{secphysbasis} by a
detailed analysis of the physical basis for this particular functional
form. In Section~\ref{secderived}, we then compute a number of derived
quantities based on the star formation history.  We briefly discuss
the expected dependence on cosmological parameters and possible
effects of metal enrichment in Section~\ref{seccosmology}, and,
finally, we summarise and conclude in Section~\ref{secconclusions}.

\section{An analytic fit to the cosmic star formation history} \label{secfit}

In \citet{SprHerSFR}, we used a set of numerical simulations to
study the evolution of the cosmic star formation rate density from
high redshift to the present.  To summarise our results compactly, we
fitted $\dot\rho_\star (z)$ using a simple double-exponential function
of the form \be \dot\rho_\star(z)= \epsilon_\star\,
\frac{b\exp\left[a(z-z_m)\right]} {b-a+a\exp\left[b(z-z_m)\right]},
\label{eqnoldfit}\ee
with $a= 3/5$, $b=14/15$, $z_m=5.4$, and $\epsilon_\star= 0.15\,{\rm
M}_\odot{\rm yr}^{-1}{\rm Mpc}^{-3}$.  (Our notation here differs
slightly from Springel \& Hernquist 2002b to avoid confusion with what
follows.)  This functional form was chosen based purely on the
suggestive shape of our numerical result.  However, we also suspected
that there should be a clear physical basis for the shape of the star
formation history, which we did not address.  With such a basis, it
should be possible to arrive at an appropriate analytic fitting
function directly, making it unnecessary to ``guess'' a particular
form for it.

\begin{figure}
\begin{center}
\resizebox{8cm}{!}{\includegraphics{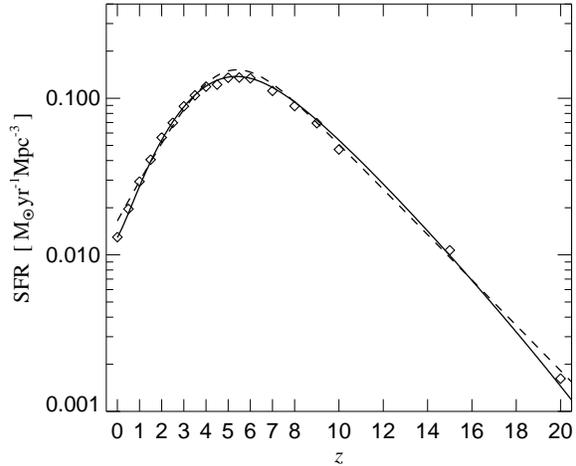}}\\%
\end{center}
\caption{\label{figanalyticfitoldnew} Simulation results for the
cosmic star formation history (symbols) compared to different analytic
fitting functions. The solid line shows equation~(\ref{eqnnewfit}),
while the dashed line gives equation~(\ref{eqnoldfit}), the fitting
function originally proposed by \citet{SprHerSFR}.}
\end{figure}

As \citet{Wh91} have demonstrated, the low-redshift behaviour of the
star formation history should be related to the declining efficiency
of gas cooling at low redshift, which in itself is caused by the
decrease of the mean density of the universe.  This effect should
hence give rise to a scaling that is related to the expansion rate of
the Universe, as measured by the evolution of the Hubble constant.
Indeed, for low redshifts, we find empirically that a dependence of
the form $\dot\rho_\star\sim H(z)^{4/3}$ matches our measurements for
our model of star formation and feedback very well.

On the other hand, at high redshifts, we clearly see a trend that is
close to an exponential.  In fact, $\dot\rho_\star\sim \exp(-z/3)$
provides an acceptable fit to our simulation results, at least over
the limited redshift range probed by our calculations. This
exponential behaviour is presumably related to the growth of the halo
mass function, which exhibits an exponential cut-off for large masses.

Based on a more thorough study of these ideas, which will be discussed
in detail below, an improved fit of the form \be \dot\rho_\star =
\dot\rho_\star(0)
\frac{\chi^2}{1+\alpha(\chi-1)^3\exp{(\beta\chi^{7/4})}}
\label{eqnnewfit}
\ee can be obtained.  Here, the redshift evolution of $\dot\rho_\star$
is conveniently captured by defining the abbreviation \be \chi(z)
\equiv \left[\frac{H(z)}{H_0}\right]^{\frac{2}{3}}, \label{eqchi} \ee
and where $\alpha=0.012$, $\beta=0.041$, and
$\dot\rho_\star(0)=0.013\,{\rm M_\odot yr^{-1}Mpc^{-1}}$ are
introduced as fitting parameters.  We find that equation
(\ref{eqnnewfit}) provides an excellent fit to our simulations, and,
in particular, is better than equation (\ref{eqnoldfit}).  This is
seen in Figure~\ref{figanalyticfitoldnew}, where we compare equations
(\ref{eqnoldfit}) and (\ref{eqnnewfit}) to our composite simulation
result.

At low redshift, we see that equation (\ref{eqnnewfit}) reduces to
$\dot\rho_\star\propto H(z)^{4/3}$, while the origin of the
high-redshift scaling $\dot\rho_\star \propto \chi^{-1} \exp(- \beta
\chi^{7/4})$ that we adopted in our fitting function, is not
immediately obvious. In fact, we have chosen this form based on a
detailed analytic argument which we will present in the next section.
This will also elucidate the dependence of the shape of the star
formation history on cosmological parameters, and on the physics
of star formation and feedback.

\section{Physical basis for the cosmic star formation history}  \label{secphysbasis}

\subsection{Basic equations}

Provided that star formation occurs only in dark matter halos, we can
compute the cosmic star formation rate density as an integral over the
multiplicity function of halos, $g(M,z)$, multiplied by the average
normalised star formation rate $s(M,z)=\left<\dot M_\star\right>/M$ of
halos of a given mass $M$.  This can be written as \be \dot\rho_\star
(z)= \overline{\rho}_0 \int g(M,z)\, s(M,z)\,{\rm dlog} M, \ee where
we term the integrand the ``multiplicity function of star formation''
\citep{SprHerSFR} and where $\overline{\rho}_0\equiv 3 \Omega_0
H_0^2/(8 \pi G)$.

The halo multiplicity function $g(M,z)$ can be defined as \be g(M,z)=
\frac{{\rm d}F}{{\rm dlog} M}, \ee where $F(M,z)$ is the fraction of
mass bound in halos less massive than $M$.  Often, $F(M,z)$ is
approximated by the \citet{Pre74} mass function, which is known to
provide a reasonable parameterisation of the evolution of halo
abundance in CDM cosmologies. The Press-Schechter mass function can be
written as \be F(M,z)= {\rm erf} \left[
\frac{\delta_c}{\sqrt{2}\,\sigma(M,z)} \right], \ee where the function
$\sigma(M,z)$ describes the linearly extrapolated rms-fluctuations in
top-hat spheres of size equal to an enclosed background mass $M$.  For
the threshold parameter $\delta_c$, we adopt the canonical value
$\delta_c=1.686$.

Recent studies have shown that there are slight deviations between the
Press-Schechter mass function with the results of high-resolution
collisionless simulations of structure formation, particularly at high
mass-scales, and around the exponential turn-off.  However,
\citet{ShTo99,ShTo02} have derived an improved parameterisation of the
mass function by generalising the excursion set formalism to allow for
ellipsoidal collapse.  We can rewrite their result in an integrated
form as \be F(M,z)= A \left[ {\rm erf} \left( {\frac{\sqrt{a}
\delta_c}{\sqrt{2}\sigma}} \right) + \frac{1}{\sqrt{2^{3/5} \pi}}
\tilde\Gamma\left(\frac{1}{5}, \frac{a\delta_c^2}{2\sigma^2}\right)
\right], \ee where $a=0.707$, $A=[1+\Gamma(1/5)/\sqrt{
2^{3/5}\pi}]^{-1}=0.3222$, and $\tilde\Gamma$ is the lower incomplete
gamma function, \be \tilde\Gamma(a,x)=\int_0^x t^{a-1}\exp(-t)\,{\rm
d}t.  \ee The Sheth \& Tormen mass function has been tested over a
large dynamic range in mass and provides an accurate description of
numerical results (Jenkins et al. 2001).  In what follows, we prefer
the Sheth \& Tormen mass function for this reason, but will also
employ the Press-Schechter form for comparison and because it works
very well at high redshift \citep{Jang01}.

The evolution of $\sigma(M,z)$ determines the evolution of the mass
function. In linear theory, we have \be \sigma^2(M,z)= D^2(z)
\int_0^{\infty} \frac{{\rm d}k}{2\pi^2} k^2 P(k)
\left[\frac{3j_1(kR)}{kR}\right]^2, \label{eqsigmadef}\ee where $D(z)$
is the linear growth factor, normalised to unity at the present time,
and $P(k)$ is the linear power spectrum. The growth factor $D(z)$ can
be computed from \be D(z)= D_0 H(z) \int_z^{\infty}\frac{(1+z'){\rm
d}z'}{H^3(z')}, \ee using the Hubble constant
\begin{eqnarray}
H(z)=H_0 \left[ \Omega_m (1+z)^3 + (1-\Omega_m-\Omega_\Lambda)(1+z)^2
  + \Omega_\Lambda\right]^{1/2} \nonumber
\end{eqnarray}
and adjusting the normalisation constant $D_0$ such that $D(0)=1$.

For the purposes of this analysis, we define halos of virial mass $M$
to be spheres of radius $R$ that enclose a characteristic overdensity
of 200 with respect to the {\em critical density}. For each halo, we
define a virial velocity \be V_{\rm vir}^2 \equiv \frac{G M}{R}. \ee
We can then express the mass and virial radius as \be M=\frac{V_{\rm
vir}^3}{10GH(z)},\;\;\;\;\;R=\frac{V_{\rm vir}}{10H(z)}. \ee We
further define the halo's virial temperature as \be T_{\rm
vir}=\frac{\mu}{2k} V_{\rm vir}^2 \simeq 36\,{\rm K}\, \left(\frac{
V_{\rm vir}}{{\rm km\,s^{-1}}}\right)^{2}, \ee where $\mu\simeq 0.6\,
m_{\rm p}$ is the mean molecular weight.  Note that $T_{\rm vir}$ is a
function only of circular velocity.  The virial temperature of a halo
of given mass $M$ at redshift $z$ is hence given by \be T(M,z)=
9.5\times 10^7 \,{\rm K}\left(\frac{M}{10^{15}\,h^{-1}{\rm
M}_\odot}\right)^{\frac{2}{3}} \, \chi(z)\, ,
\label{eqntvir}
\ee
where $\chi(z)$ is defined in equation (\ref{eqchi}).

\subsection{A model for the star formation efficiency}

Given that the Sheth \& Tormen mass function specifies $g(M,z)$
unambiguously, it is clear that the key for an explanation of the full
star formation history lies in an understanding of the evolution of
the normalised star formation rate $s(M,z)$.  In \citet{SprHerSFR}, we
measured $s(M,z)$ directly at different epochs from our set of
hydrodynamic simulations.  We found that only halos with virial
temperatures above $\simeq 10^4\,{\rm K}$ formed any stars, which is
simply caused by the inefficiency of atomic line cooling at lower
temperatures, when metals and molecular cooling are neglected.  While
molecular cooling might be of high importance for the formation of the
very first stars in the high-redshift universe, it should be largely
unimportant for the formation of the bulk of the stars. On the other
hand, metal line cooling may boost the cooling rates in halos at late
times, provided their diffuse gas becomes significantly enriched with heavy
elements. In section~\ref{secmetalcooling}, we will discuss
separately to what extent our neglect of metal cooling could influence
our results.

From our simulations, we further found that the normalised star
formation rate, expressed as a function of virial temperature, has
approximately the same shape at different redshifts, differing only in
amplitude.  This can be expressed formally by defining a function \be
\tilde s(T)\equiv s[M(T,z),z], \;\; {\rm at}\;\; $z=0$, \ee where
$M(T,z)$ is the mass of a halo of virial temperature $T$ at redshift
$z$.  The inference of the near shape invariance of $s(M,z)$ when
expressed as a function of virial temperature then allows us to make
the ansatz \be s(M,z)=\tilde s[T(M,z)]q(z), \label{eqsansatz}\ee where
$q(z)$ describes the scaling of the normalised star formation rate
with redshift, and $T(M,z)$ is given by equation (\ref{eqntvir}).  We
note that in different models for the physics of star formation
and feedback it may not be possible to factorise the star formation
rate in the form of equation~(\ref{eqsansatz}). However, we expect
this ansatz to work for models that are broadly similar to the one
considered in our simulations.

Schematically, $\tilde s(T)$ vanishes for temperatures below
$10^4\,{\rm K}$.  For higher temperatures, it assumes a roughly
constant value of $\tilde s(T) \simeq s_0$, up to $T\approx
10^{6}\,{\rm K}$, where it begins to rise by about a factor of 3
towards a maximum reached around $10^7\,{\rm K}$, beyond which $\tilde
s(T)$ starts to fall again towards higher temperatures.  The detailed
shape of $\tilde s(T)$ is in part related to the strong feedback by
galactic winds considered in our simulations.  These winds are an
important mechanism for maintaining the normalised star formation rate
at a relatively low level in small halos within the temperature range
$10^4-10^6\,{\rm K}$. Likewise, the scale at which the normalised star
formation rate starts to increase is related to the speed of the
winds.  When they are no longer able to escape from halos, the winds
loose their ability to suppress star formation.

We note that perhaps one of the most crucial characteristics of
$\tilde s(T)$ is its {\em threshold behaviour}: Below a critical
temperature of $T_4\equiv 10^4\,{\rm K}$, there is no star formation,
while above $T_4$, the value of $\tilde s(T)$ varies only relatively
weakly with temperature.  Perhaps the simplest reasonable model for
the normalised star formation rate therefore takes the form of a
step-function: \be \label{eqnqsimple} s(M,z)=\tilde s(T) q(z)=\left\{
\begin{array}{cl}
s_0\, q(z) & {\rm for}\;\;T> 10^4\,{\rm K}, \\
0 &  {\rm otherwise}.\\
\end{array}
\right.  \ee

\begin{figure}
\begin{center}
\resizebox{8cm}{!}{\includegraphics{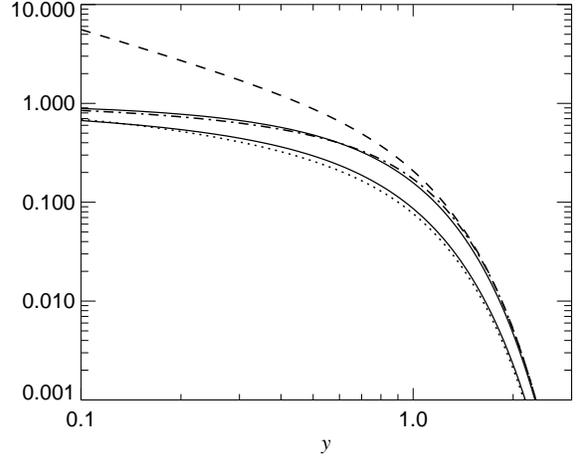}}\\%
\end{center}
\caption{\label{figerrfunc} Different approximations for the mass
fraction above star formation threshold. The upper solid line shows
the exact value of $1-{\rm erf}(y)$. The dashed line is the
approximation based on equation~(\ref{eqx}), while the dot-dashed line
shows the approximation~(\ref{eqxx}). Similarly, the lower solid line
and the dotted line compare the corresponding exact expression for the
Sheth \& Tormen mass function with our approximation, respectively, as
given in equation~(\ref{eqnstappr}).}
\end{figure}

If we define $M_4(z)$ to be the mass corresponding to a virial
temperature of $10^4\,{\rm K}$ at redshift $z$, this model immediately
implies \be \dot\rho_\star (z) = \overline{\rho}_0 \, s_0\, q(z)
\left[ F(\infty, z) - F(M_4, z)\right] .  \ee Using the
Press-Schechter mass function, this becomes \be \dot\rho_\star (z) =
\overline{\rho}_0 \, s_0\,q(z) \left[ 1 - {\rm erf}\left(
\frac{\delta_c}{\sqrt{2}\,\sigma_4} \right) \right],
\label{eqnsfrPS}\ee where we have introduced
$\sigma_4(z)\equiv\sigma[M_4(z),z]$ to describe the rms-fluctuations
on the mass scale of the $10^4\,{\rm K}$ halos.  If instead the Sheth
\& Tormen mass function is used, we obtain
\begin{eqnarray} \dot\rho_\star (z) & =&
\overline{\rho}_0\, s_0\,q(z)  \label{eqnsfrST} \\
& &  \left[ 1 - A\,{\rm erf}\left(
\frac{\sqrt{a}\delta_c}{\sqrt{2}\,\sigma_4}\right)
-\frac{A}{\sqrt{2^{3/5} \pi}} \tilde\Gamma\left(\frac{1}{5},
\frac{a\delta_c^2}{2\sigma_4^2}\right) \right].
\nonumber
\end{eqnarray}

At high redshift, we expect the arguments of the error functions in
equations (\ref{eqnsfrPS}) and (\ref{eqnsfrST}), respectively, to be
large compared to unity. We can then use an asymptotic expansion of
the error function \citep[e.g.][]{Gradstein81} to obtain simpler
approximate expressions.  To lowest order we have \be 1-{\rm
erf}(y)\simeq \frac{1}{\sqrt{\pi}} \, \frac{\exp(-y^2)}{y} .
\label{eqx} \ee 
This approximation is very accurate for $y\gg1$, and is even 
reasonable for $y\sim 1$.
For $y>2$, the error is less than
10\%, but it grows to 30\% for $y\simeq 1$, reaching a factor of 2 for
$y\simeq 0.45$.  However, because the values of $y$ we encounter in
equations (\ref{eqnsfrPS}) and (\ref{eqnsfrST}), respectively, drop to
about 0.2 for $z=0$, we desire a more accurate approximation at low
$z$.  In fact, we propose that \be 1-{\rm erf}(y)\simeq \frac{1}{1 +
\sqrt{\pi} \, y \,{\exp(y^2)}}\label{eqxx} \ee fullfills our
requirements very well.  This approximation is accurate to better than
12\% for all $y\ge 0$.  In Figure~\ref{figerrfunc}, we compare the
approximations~(\ref{eqx}) and (\ref{eqxx}) to the exact result.  Also
shown is the relevant expression for the Sheth \& Tormen mass
function, where we find that the approximation \be 1-A\,{\rm
erf}(y)-\frac{A}{\sqrt{2^{3/5}\pi}}\tilde\Gamma(\frac{1}{5},y^2)
\simeq \frac{1}{1 + \frac{5}{2}\sqrt{\pi} \, y \,{\exp(y^2)}}
\label{eqnstappr} \ee provides a similarly small
error.

At high redshift, when $\sigma_4(z)$ is small, we hence expect the
star formation rate from the Sheth \& Tormen mass function to scale as
\be \dot\rho_\star (z) \propto q(z)\sigma_4(z)
\exp\left[-\frac{a\delta_c^2}{2\sigma_4^2}\right]. \ee In the case of
the Press-Schechter form, the numerical factor $a=0.707$ in the argument of
the exponential function would be absent, giving a somewhat faster
decline towards high redshift. To understand how fast this suppression
develops with redshift, we need to understand the scaling of
$\sigma_4(z)$ and $q(z)$ separately.

\begin{figure}
\begin{center}
\resizebox{7.5cm}{!}{\includegraphics{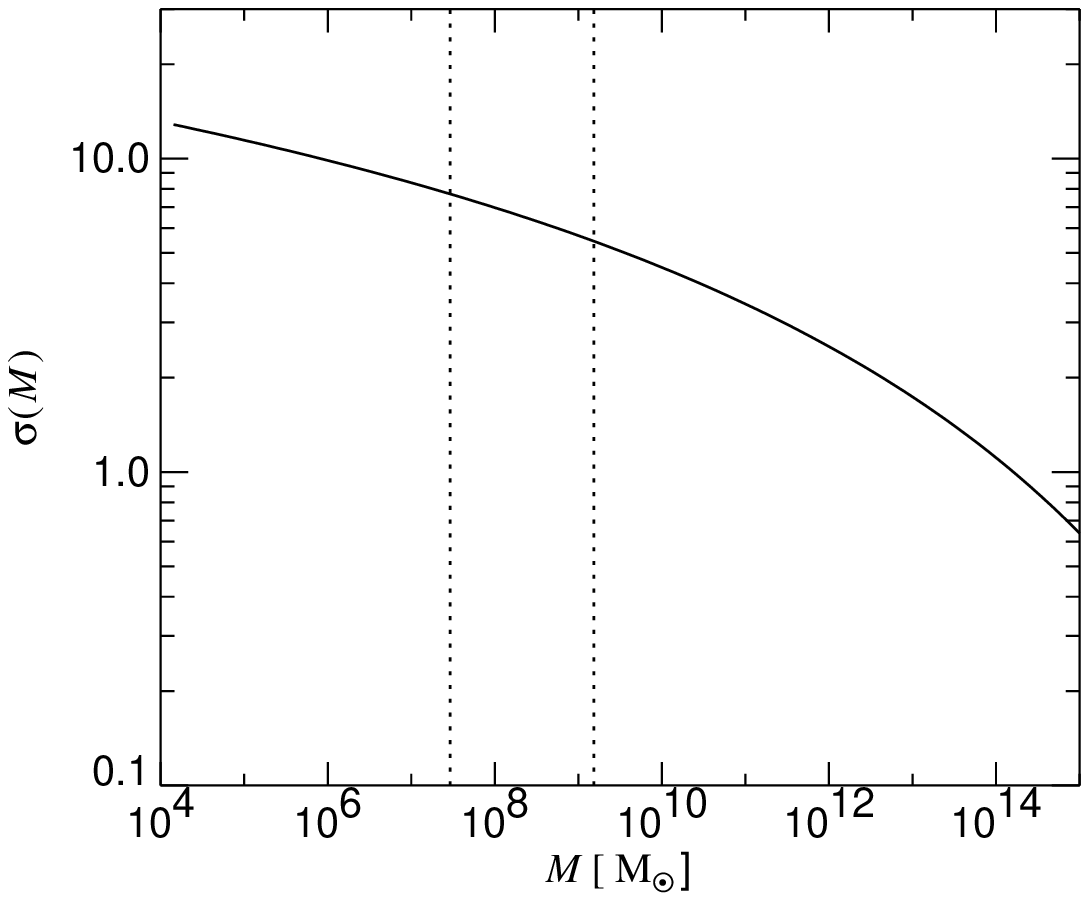}}\\%
\resizebox{7.5cm}{!}{\includegraphics{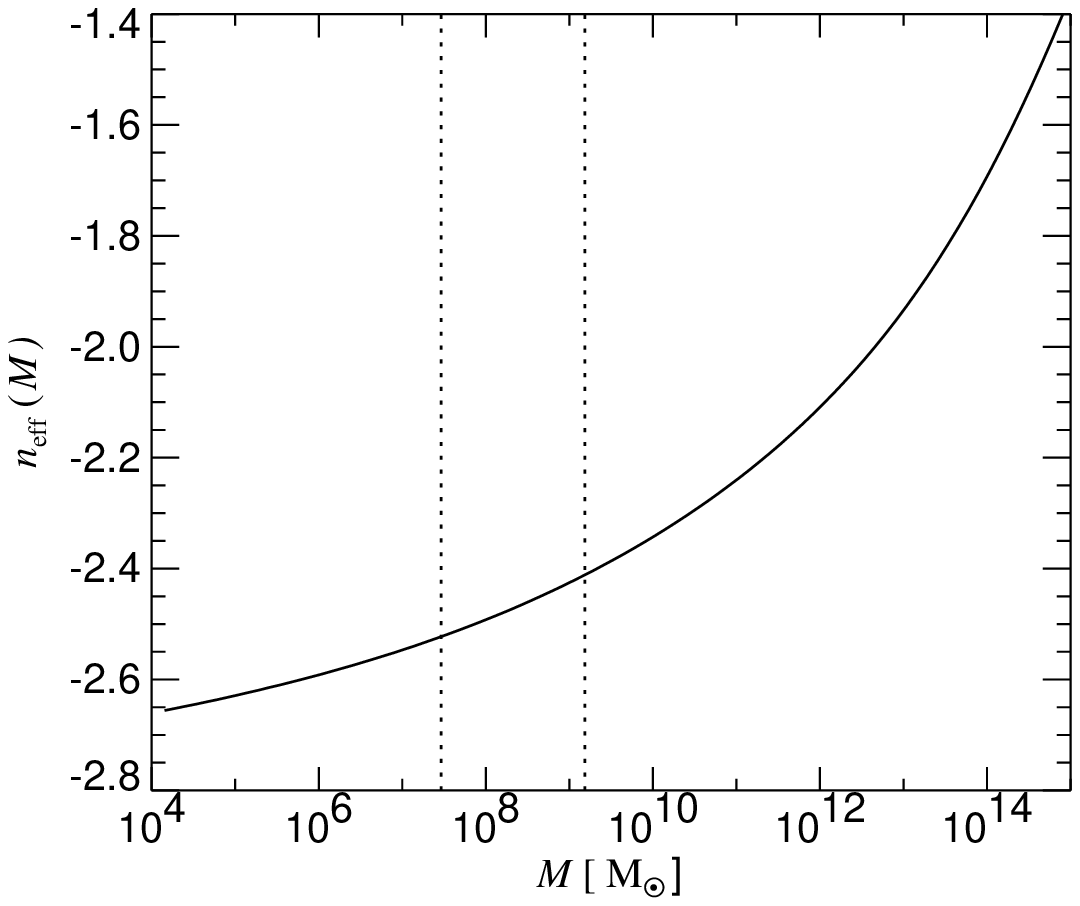}}\\%
\resizebox{7.5cm}{!}{\includegraphics{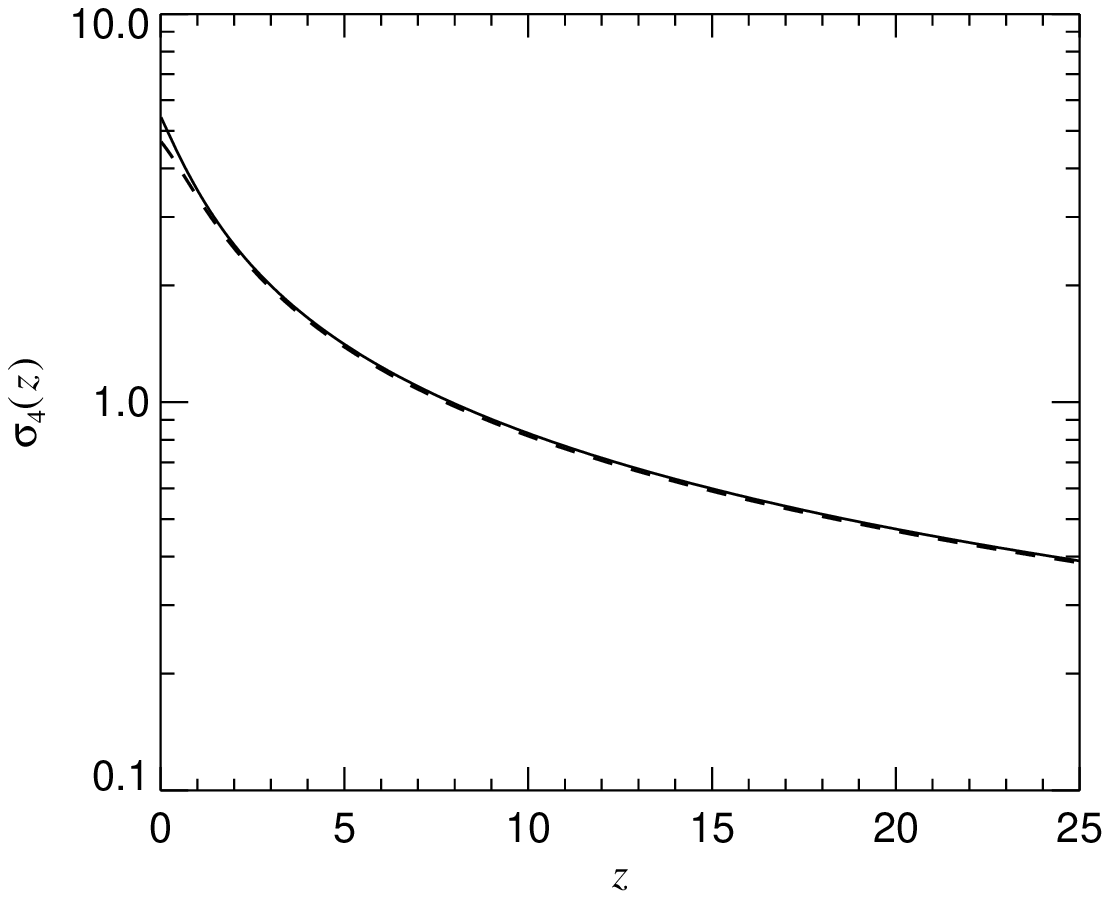}}\\%
\end{center}
\caption{\label{figs4} Top panel shows the rms-fluctuations
$\sigma(M)$ in top-hat spheres that enclose a background mass $M$, for
the present day linear power spectrum of a $\Lambda$CDM cosmology.  In
the middle panel, we show the corresponding effective slope of the
power spectrum, defined here as $n_{\rm eff}= -3 -6\, {\rm
dlog}\sigma/{\rm dlog}M$. The dotted vertical lines indicate the range
of masses that correspond to a virial temperature of $10^4\,{\rm K}$
between $z=20$ (left line) and $z=0$ (right line). In the bottom
panel, we show the evolution of $\sigma_4(z)$ for the $\Lambda$CDM
cosmology. The dashed line shows the approximation of
equation~(\ref{eqnsigma4approx}).}
\end{figure}

\subsection{The scaling of $\sigma_4(z)$}

If we approximate the power spectrum on the scales of interest by a
power law, $P(k)\propto k^n$, then the density fluctuations scale as
$\sigma^2(M)\propto M^{-(n+3)/3}$ at fixed redshift.  Based on
equation~(\ref{eqsigmadef}), we hence have \be \sigma^2_4(z) \propto
D^2(z) [M_4(z)]^{-\frac{n+3}{3}} \propto D^2(z)
[\chi(z)]^{\frac{n+3}{2}}, \label{eqs4a}\ee where in the last step we
made use of the conversion between mass and temperature defined in
equation~(\ref{eqntvir}), and we used our definition of $\chi(z)$
given in equation~(\ref{eqchi}).

Let us now consider the growth factor. Restricting ourselves to
spatially flat cosmologies, we can write it as \be D(z)=D_0
H_0^{-2}\,\Omega_0^{1/3} \chi^{{3}/{2}}(z) \int_{\chi(z)}^{\infty}
\frac{{\rm d}y}{y^{{7}/{2}}}
\left(1-\frac{\Omega_\Lambda}{y^3}\right)^{-\frac{1}{3}}. \ee Because
we have $\Omega_\Lambda/y^3<1$, we can expand the integrand in a
Taylor series and integrate term by term. This gives \be D(z)= D_0
H_0^{-2}\,\Omega_0^{1/3} \chi^{-1}
\left[1+\frac{5}{33}\frac{\Omega_\Lambda}{\chi^3}+{\cal
O}(\chi^{-6})\right]. \ee  Therefore, to lowest order we have \be D(z)
\propto \frac{1}{\chi(z)}, \ee which is accurate to better than 15\%
for the $\Lambda$CDM cosmology.  Combining this with the result
obtained in equation~(\ref{eqs4a}), we then arrive at the expected
scaling of $\sigma_4(z)$ in the form of \be \sigma_4(z) \propto
\chi^\frac{n-1}{4}. \ee

In Figure~\ref{figs4}, we show the rms-fluctuations $\sigma(M)$ for
the linear present-day power spectrum of a $\Lambda$CDM cosmology,
together with its effective slope.  On the mass scales that correspond
to a virial temperature of $10^4\,{\rm K}$, we find that $n$ varies in
the range $-2.53$ to $-2.43$ between $z=20$ and $z=0$, so that we can
approximate it with $n\simeq -2.5$. In the
bottom panel of Figure~\ref{figs4}, we compare an exact computation of
$\sigma_4(z)$ with our predicted scaling of \be \sigma_4(z) = 4.7
\chi^\frac{n-1}{4} ,
\label{eqnsigma4approx}\ee using this effective slope of $n= -2.5$.
The maximum error is $\simeq 13\%$ and occurs at the present epoch,
where our approximation of the growth factor starts to loose
precision, but this accuracy is still sufficient for our purposes.

\subsection{The scaling of $q(z)$}

Above, we have made the approximation that the evolution of the halo
mass function depends only on gravitational physics, and as such can
be described by the well-established results for non-linear
structure formation in cold dark matter universes.  It is clear,
however, that the evolution of the star formation efficiency involves
more complicated baryonic processes as well. Unfortunately, the
relevant physics of radiative cooling, star formation, and feedback,
is much less well understood.

In the simulations presented in \citet{SprHerSFR}, we used a
novel parameterisation of star formation and feedback in terms of a
sub-resolution model of a two-phase interstellar medium
\citep{SprHerMultiPhase}. In addition, we included strong galactic
winds as a phenomenological extension of the model. Despite the
complexity of these physical processes, we found that the
normalised rate of star formation followed the simple factorisation
suggested in equation~(\ref{eqsansatz}); i.e.~the shape of $s(M,z)$
remains approximately invariant when expressed as a function of
virial temperature, while the amplitude of $s(M,z)$ scales with
redshift.

Our results indicated that the normalised star formation rate scales
steeply with redshift, roughly as $\sim (1+z)^3$, over the redshift
range $2<z<7$.  At redshifts below about $z\sim 2$, this evolution
clearly appeared to slow down, however. At very high redshifts, for
$z>7$, the scaling also became much slower, apparently becoming
approximately constant towards even higher redshifts.

We here argue that in the context of our model for star formation and
feedback this behaviour can be understood in terms of two effects:
\begin{enumerate}
\item At low and intermediate redshifts, cooling is relatively slow,
such that one can ultimately expect that, at fixed virial temperature,
the star formation rate scales in proportion to the cooling rate
of a halo. For very massive halos, this is evident, because due to the
inefficiency of feedback in massive halos, the supply of fresh cold
gas directly governs the star formation rate.  For smaller halos,
feedback processes make star formation less efficient than the cooling
rate, but the resulting net amplitude can still be expected to vary in
proportion to the cooling rate of the halo, provided that the dynamical
equilibrium between star formation, cooling and feedback responds
linearly to variations in the cooling rate.
\item At very high redshifts, cooling is rapid because of the high
mean density of halos, but the strength of feedback processes in halos
of fixed virial temperature remains unchanged. We should then
encounter a regime where the star formation rate is no longer determined
by the cooling rate, but instead by the gas consumption timescale
$t_0^\star$ used in our multi-phase model of star formation
\citep[see][]{SprHerMultiPhase}.  In this regime, we can picture the
gas within halos to be cooling so rapidly that it essentially all becomes
cold instantly, so that the star formation rate would asymptote
to something of order $\sim M_{\rm cold}/\left<t^\star\right>$, with
$M_{\rm cold}$ being roughly equal to the total gas mass in the halo,
and $\left<t^\star\right>$ being of order $t_0^\star$ (slightly
smaller probably because of the decline of the consumption timescale
towards higher density).
\end{enumerate}

We now try to make this picture more quantitative.  For the first
point, we need an estimate of the cooling rate and how it scales with
redshift.  In order to obtain this, we use a variant of the cooling
model employed in \citet{Spr99b} and \citet{Yoshida2001}, which in
itself is similar to the model of \citet{Wh91}.

The model starts by assuming that the hot gas is distributed according
to a spherically symmetric profile $\rho_g(r)$ within a halo, with the
gas being at the halo virial temperature $T$. We define a local
cooling time by \be t_{\rm cool}(r)= \frac{3\, k T
\,\rho_g(r)}{2\,\mu\,n_{\rm H}^2(r)\Lambda(T) }, \label{cooltime} \ee
where $n_{\rm H}(r)$ is the number density of hydrogen, $\mu$ the
molecular weight, and $\Lambda(T)$ the cooling function.

If the density profile is assumed to remain approximately fixed during
cooling, the gas in the halo will have cooled at time
$t$ out to a radius $r_{\rm cool}(t)$ given by \be t_{\rm cool}[r_{\rm
cool}(t)]= t\label{eqxyz}. \ee This allows the cooling rate to be
estimated as \be \frac{{\rm d}M_{\rm cool}}{{\rm d}t} = 4\pi
\rho_g(r_{\rm cool})\, r_{\rm cool}^2 \,\frac{ {\rm d} r_{\rm cool}} {
{\rm d}t}.  \ee Most of the cooling models used in semi-analytic
calculations of galaxy formation are based on this equation, but they
vary in the assumptions made about the profile $\rho_g(r)$, and the
perhaps more uncertain question as to what time $t$ should be
used in equation (\ref{eqxyz}). For example, \cite{Wh91} have proposed
using the age of the universe for $t$. It has also been
argued that $t$ should be the time elapsed since the last major merger
of a halo \citep{SomPr99}. \citet{Spr99b} suggest using the dynamical
time of the halo instead, arguing that the gas profile should
react to pressure losses from cooling on this timescale, and hence the
cooling radius can on average be expected to grow to a radius
corresponding to this time.

None of these choices can be rigorously justified without treating the
dynamics of the gas self-consistently, which is beyond the scope of a
simple analytic estimate.  However, \citet{Yoshida2001} have directly
compared cooling rates measured in hydrodynamic simulations with the
above semi-analytic cooling model and find quite good agreement for a
parameterisation where the ansatz with the dynamical time was
used.  We will therefore choose it in what follows.

For the gas profile, a truncated isothermal sphere with
$\rho(r)\propto r^{-2}$ is often adopted. To allow the possibility
that the gas profile may have a different slope than an isothermal
one at the typical cooling radius, we write the density profile in a
slightly more general form as \be \rho_g(r)=\frac{(3-\eta)\,M_{\rm
gas}} {4\pi\,R_{\rm vir}^{3-\eta}\,r^\eta}, \ee so
that the profile behaves as $\rho(r)\propto r^{-\eta}$ at the
typical cooling radius.  From equation (\ref{eqxyz}) we find \be
\frac{{\rm d} r_{\rm cool}} {{\rm d}t }= \frac{1}{\eta}\frac{r_{\rm
cool}}{t_{\rm cool}}. \ee Noting that we set $t_{\rm cool}=t_{\rm
dyn}\equiv R_{\rm vir}/V_{\rm vir}$, we then obtain \be r_{\rm cool}=
\left[ \frac{ (3-\eta) \,M_{\rm gas} R_{\rm vir}^{\eta-2}} {4\pi f(T)
V_{\rm vir}}\right]^{\frac{1}{\eta}}, \ee where we defined the
abbreviation \be f(T)= t_{\rm cool}(r) \rho_g(r) =\frac{3 m_{H}^2 k
T}{2\mu X^2\Lambda(T)}, \ee and $X$ is the hydrogen mass fraction.
The cooling rate follows as \be \frac{{\rm d}M_{\rm cool}}{{\rm d}t}
=\frac{3-\eta}{\eta} f_b M_{\rm vir} \left[\frac{(3-\eta)f_b}{4\pi G
f(T)}\right]^{\frac{3-\eta}{\eta}} \,\left[ 10
H(z)\right]^{\frac{3}{\eta}}, \label{eqncoolrate}\ee with $f_b\equiv
M_{\rm gas} / M_{\rm vir}$.

We can now quantitatively check how well this estimate of the cooling
rate explains the values of the star formation rate we measured in our
simulations for large halos at late times.  Recall that we argued
that, for halos large enough to be unaffected by feedback effects, the
star formation rate should essentially be given by the cooling
rate. This should then clearly be the case for a virial temperature of
$10^7\,{\rm K}$, for example.  This is because the galactic winds of
our simulations, which leave star forming regions at a velocity of up
to $v_{\rm wind}=484\,{\rm km}\,{\rm s}^{-1}$, can at most heat gas up
to temperatures of $\sim 5\times 10^6\,{\rm K}$ when they are stopped,
and they are expected to be unable to escape the gravitational
potential well of halos with virial temperatures significantly larger
than $\sim 10^6\,{\rm K}$.

\begin{figure}
\begin{center}
\resizebox{8.0cm}{!}{\includegraphics{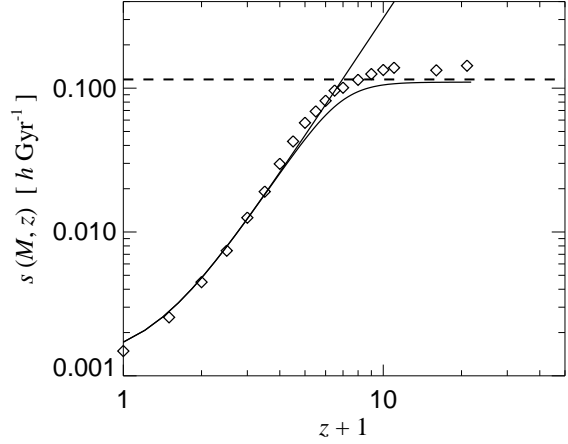}}%
\caption{\label{figcomp} Comparison of the measurement of $s(M,z)$
from the simulations (symbols) at a fixed virial temperature of
$T=10^7\,{\rm K}$ with estimates based in the cooling rate (solid),
and on the ``limit'' argument (dashed).}
\end{center}
\end{figure}

In Figure~\ref{figcomp}, we show the measurements of $s(M,z)$ we made
at a virial temperature of $10^7\,{\rm K}$.  We compare this data to
the cooling rate predicted by equation (\ref{eqncoolrate}), noting that the
cooling function has a value of \be \Lambda(T)/n_{\rm H}^2 \simeq
10^{-23}\, {\rm erg}\,{\rm cm}^3\,{\rm s}^{-1} \ee at $T= 10^7\,{\rm
K}$.  We expect that we should find roughly $s\simeq \dot M_{\rm
cool}/{M_{\rm vir}}$, and in fact, an almost perfect fit is obtained
for $\eta=1.65$ out to redshift $z\sim7$. Given that the cooling model
is rather crude, this level of agreement is remarkable.  On the other hand,
because the cooling model is so simple, one should probably take the
good fit for $\eta=1.65$ with a grain of salt, and grant that a different
value in the range $\eta=1.5-2$ may also be acceptable. Note, however,
that $\eta=1.65$ does quite a bit better in reproducing the shape
traced out by the measurements than the isothermal value $\eta=2.0$,
which would predict a result lying a bit above the measurements. Also
note that for $\eta\simeq 1.5$, the model predicts a scaling $\propto
H(z)^2$, which is essentially equal to the $\propto(1+z)^3$ scaling
that we had guessed empirically at intermediate redshift.

However, at high redshift, the measurements for $s(M,z)$ clearly fall
short of this scaling and instead appear to approach some kind of
limit.  This brings us back to the second point discussed above, where
we argued that the model used to describe star formation implicitly
imposes a maximum on the star formation rate in a given halo. If
essentially all the gas in a halo has cooled, we expect of order $f_b
M_{\rm vir}$ of cold gas, where $f_b=\Omega_b/\Omega_0=0.133$ is the
universal baryon mass fraction. In our model for star formation, a
fraction $x\sim 0.95$ of this cold gas is in clouds, such that the
maximum star formation rate can be estimated as $\dot M_\star \sim
(1-\beta)xf_b M_{\rm vir}/\left<t^{\star}\right>$, where $\beta=0.1$
is the mass fraction of short-lived stars. The typical star formation
timescale $\left<t^{\star}\right>$ of the gas will be somewhat smaller
than the parameter $t_0^\star$ used in our star formation model,
because of the density dependence of the consumption timescale. If we
roughly estimate $\left<t^{\star}\right>= 2/3\, t_0$, then we obtain
$\dot M_\star/M_{\rm vir} \simeq 0.12\,h\,{\rm Gyr}^{-1}$, in quite
good agreement with the suggested maximum value based upon the
measurements, as seen in Figure~\ref{figcomp}.

We can incorporate this maximum value into our expected scaling of \be
q(z)= \chi^{\frac{9}{2\eta}} \label{eqq1}\ee based on the cooling rate
alone, by making the replacement \be \chi \rightarrow \frac{\chi
\tilde\chi}{(\chi^m+\tilde\chi^m)^{\frac{1}{m}}}, \label{eqq2}\ee
where $\tilde\chi$ is a constant that limits $\chi(z)$ at high
redshift.  This functional form provides a smooth transition between
the regime that is governed by $q(z)= \chi^{\frac{9}{2\eta}}$, and the
one where $q(z)$ becomes constant.  For larger values of $m$, the
transition can be made sharper. We obtain a good match to our
simulation results for $\tilde\chi=4.6$ and $m=6$, as shown in
Figure~\ref{figcomp}.

In summary, the above discussion gives a plausible quantitative
explanation for the behaviour of the normalised star formation rate in
halos massive enough such that feedback effects are unimportant.
However, for halos of small virial temperature, the simple derivation
given above breaks down to some extent, because here feedback
processes and winds {\em are clearly important}. In particular, the
star formation rates measured for low mass halos in our
simulations are substantially smaller than expected based on the
cooling rate alone. At a fixed epoch, $\tilde s(T)$ assumes a value
about 3 times smaller for virial temperatures below $\sim 5\times
10^6\,{\rm K}$ than at the peak at $\simeq 10^7\,{\rm K}$.  We argue
that this behaviour is largely caused by feedback processes, notably
by galactic winds. Nevertheless, the scaling of the star formation
rate with redshift can still be described in terms of $q(z)$
given in equations~(\ref{eqq1}) and (\ref{eqq2}).

\subsection{A general fitting formula}

Summarising the above, we have derived an analytic expression for the
expected evolution of the star formation rate with redshift. Focusing
on the Sheth \& Tormen mass function in the following, which is known
to provide a very good fit to the mass function measured in
cosmological simulations, we have \be \dot\rho_\star(z) =
\overline{\rho}_0 s_0 \left[ \frac{\chi
\tilde\chi}{(\chi^m+\tilde\chi^m)^{\frac{1}{m}}}
\right]^{\frac{9}{2\eta}} \frac{1}{1 + \frac{5}{2}\sqrt{\pi} \, u
\,{\exp(u^2)}}, \label{eqnbasicmodel} \ee where $u=\sqrt{a/2}\,
\delta_c/\sigma_4(z) \simeq 0.21\chi^{7/8}$, and
$\chi(z)=[H(z)/H_0]^{2/3}$. Recall that we determined
$\tilde\chi=4.6$, $\eta=1.65$, and $m=6$ as a fit to the scaling of
the normalised star formation rate in halos of fixed virial
temperature.

\begin{figure}
\begin{center}
\resizebox{8.0cm}{!}{\includegraphics{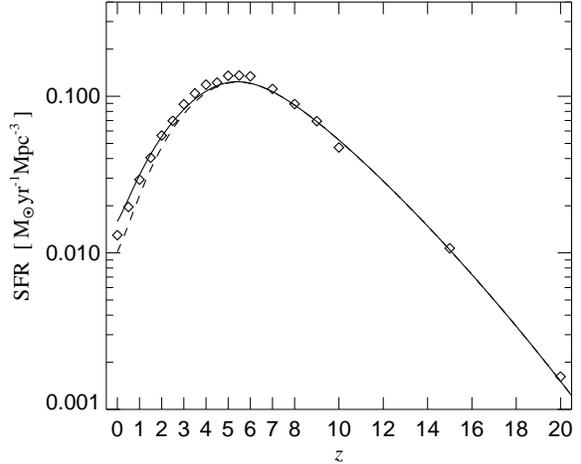}}%
\caption{\label{figfit} Comparison of simulation results for the
cosmic star formation history (symbols) with expectations based on
analytic estimates. The dashed line shows the result of
equation~(\ref{eqnbasicmodel}), where all halos with higher virial
temperature than $10^4\,{\rm K}$ were assumed to form stars with equal
efficiency. The solid line shows the model of
equation~(\ref{eqnfullresult}), where an additional contribution from
halos more massive than $10^{6.5}\,{\rm K}$ was included.}
\end{center}
\end{figure}

In Figure~\ref{figfit}, we compare this equation to the direct
simulation result. We see that the shape is indeed reproduced very
well by the fitting function. The fit is nearly perfect, except that
the ratio of high-redshift to low-redshift star formation appears not
to be fully correct yet.  When normalised to the star formation seen
at high $z$, the analytic expression predicts slightly too little star
formation at low redshift.

The reason for this lies in our very simplistic threshold model for
the variation of the normalised star formation rate with temperature,
which neglected the fact that the star formation efficiency is
actually not strictly constant for temperatures above $10^4\,{\rm K}$.
Indeed, the presence of feedback by galactic winds maintains the
normalised star formation rate roughly at a constant level for
temperatures below $\simeq 10^{6.5}\,{\rm K}$, above which it rises to
about three times higher.

We can incorporate this effect roughly into our model for the scaling
of the normalised star formation rate by replacing
equation~(\ref{eqnqsimple}) with \be \tilde s(T)\,q(z)=\left\{
\begin{array}{cl}
s_0\,q(z) & {\rm for}\;\;10^4\,{\rm K}<T<10^{6.5}\,{\rm K}, \\
3s_0\,q(z) & {\rm for}\;\; 10^{6.5}\,{\rm K} < T,\\
0 &  {\rm otherwise}.\\
\end{array}
\right.  \ee We can easily predict the star formation rate for this
ansatz using the equations derived previously. All we need
is the scaling of $\sigma_{6.5}(z)$, the rms-fluctuations in
spheres of mass-scale corresponding to a virial temperature
$10^{6.5}\,{\rm K}$.  For this, we obtain 
\be \sigma_{6.5}(z) \simeq 1.5 \chi^{
\frac{n'-1}{4}}, \ee and note that the effective slope of the power
spectrum on these mass scales is $n' \simeq-2.1$.  This
allows us to write the star formation history as \bea
\dot\rho_\star(z) = \overline{\rho}_0 s_0 \left[ \frac{\chi
\tilde\chi}{(\chi^m+\tilde\chi^m)^{\frac{1}{m}}}
\right]^{\frac{9}{2\eta}} \left[\frac{1}{1 + \frac{5}{2}\sqrt{\pi} \,
u \,{\exp(u^2)}}\, + \right.\nonumber\\ \left.  \frac{2}{1 +
\frac{5}{2}\sqrt{\pi} \, v \,{\exp(v^2)}} \right]
\label{eqnfullresult}\eea with $v$ being defined as $v=\sqrt{a/2}\,
\delta_c/\sigma_{6.5}(z) \simeq 0.67\chi^{0.78}$, based on the
scaling of $\sigma_{6.5}(z)$.

We also show this expression in Figure~\ref{figfit}, where it is seen
that it fits our simulation results very well. Also note
that the normalisation for $s_0$ we picked here is only about $\sim
10\%$ different from the value predicted by Figure~\ref{figcomp}, if
one identities the $z=0$ measurement of this $10^7\,{\rm K}$ halo with
$3s_0$.  We think that this good agreement is quite encouraging,
showing that we have correctly modelled the effects that determine the
evolution of the cosmic star formation density in our simulations.

It is interesting to consider the low and high redshift behaviour of
the above expressions separately. At high redshift, the normalised
star formation rate looses its redshift dependence, and only halos of
virial temperature $10^4\,{\rm K}$ and slightly above contribute
significantly to the star formation rate.  Furthermore, we have $u\gg
1$ in this regime, so that the star formation rate then scales as \be
\dot\rho_\star(z) \propto \chi^{\frac{n-1}{4}} \exp\left(-\beta\,
\chi^{\frac{1-n}{2}}\right), \label{eqnhighz} \ee where \be \beta=a
\delta_c^2/[2 \sigma_4^2(0)]=0.044 \label{eqnbeta},\ee and $n=-2.5$ is
the appropriate effective slope of the power spectrum. It is important
to note that this exponential decline of the star formation rate
towards high redshift is directly related to the growth of the mass
function, and has a purely gravitational origin.  It arises as a
consequence of the threshold behaviour of the star formation
rate, which is in itself caused by the properties of atomic line
cooling.  Despite this, the high-redshift behaviour of the star
formation rate is generic, and independent of details of star
formation itself. Interestingly, the decline towards high redshift
depends on the shape and normalisation of the power spectrum.

Note that the decline of $\log\dot\rho_\star$ towards high redshift is
not strictly linear in redshift, as we had suspected earlier when
deriving a first empirical fit to our simulation results. However,
over the limited redshift range $6<z<20$, the scaling of
equation~(\ref{eqnfullresult}) is well fit by a simple $\propto
\exp(-z/3)$, which we had guessed originally. But at still higher
redshift, we expect the star formation rate to decline significantly
faster than this.

At low redshift, the normalised star formation rate scales as
$q(z)\propto \chi^{\frac{9}{2\eta}}$, while the exponential growth of
the mass fraction above the star formation threshold of $10^4\,{\rm
K}$ essentially ends, being replaced by a comparatively slow residual
increase.  The combination of these two effects leads to the decline
of the star formation density at low redshift.

To examine the low redshift behaviour itself, we note that
equation~(\ref{eqnstappr}) can be further simplified, because $y$ lies
in a limited range between 0.2 and 0.8 for $0<z<5$. We can then use
the approximation \be \frac{1}{1 + \frac{5}{2}\sqrt{\pi} \, y
\,{\exp(y^2)}}\simeq\frac{1}{10\,y},\ee which is good to better than
10\% in the range $0.2<y<0.8$.  Noting that $1/y \propto \sigma_4$, we
therefore expect the low-redshift star formation rate to scale as \be
\dot\rho_\star(z) \propto \chi^{\frac{9}{2\eta} +
\frac{n-1}{4}}. \label{eqnlowz}\ee Since at very low redshift, the
additional contribution of halos more massive than $10^{6.5}\,{\rm K}$
begins to dominate, it may be more appropriate to use the effective
slope of $n'=-2.1$ in this equation instead of $n=-2.5$ for the $M_4$
mass scales.

A simple analytic form that smoothly joins the low-z
behaviour~(\ref{eqnlowz}) and the high redshift
scaling~(\ref{eqnhighz}) is given by \be \dot\rho_\star(z) =
\dot\rho_\star(0) \frac{\chi^{\frac{9}{2\eta} + \frac{n'-1}{4}}}{
1+\alpha(\chi-1)^{\frac{9}{2\eta}+\frac{n'-n}{4}} \exp\left(\beta\,
\chi^{\frac{1-n}{2}}\right)}. \ee The exponent of $\chi$ in the
numerator can be approximated as ${9}/({2\eta}) +
({n'-1})/{4}=1.95\simeq 2$. Similarly, the exponent of $\chi$ in the
denominator is approximately $9/(2\eta)+(n'-n)/4=2.83\simeq 3$, while
the exponent in the argument of the exponential function is $(1-n)/2=
7/4$.  We can hence consider a simplified fitting function of the form
\be \dot\rho_\star(z) = \dot\rho_\star(0) \frac{\chi^2}{
1+\alpha(\chi-1)^{3} \exp\left(\beta\, \chi^{{7/4}}\right)}. \ee This
is the expression we proposed at the onset of our
analysis.  Compared to our full analytic estimate of the evolutionary
history, it has the advantage of a simpler analytic form, but
involves a fitting parameter $\alpha$. For $\alpha=0.012$ and
$\beta=0.041$, we obtain a very good fit to our simulation result, as
seen in Figure~\ref{figanalyticfitoldnew}. Note that $\beta$ is in
principle determined by equation~(\ref{eqnbeta}), and thus depends
directly on the normalisation of the power spectrum. The reason why we
here lowered $\beta$ slightly from $0.044$ to $0.041$ was just to
approximately compensate the increase of the leading exponent of
$\chi$ in the denominator from $2.83$ to $3$ that we made.

\subsection{The peak of the star formation history}

As we have seen, with time the star formation rate falls at low
redshift, while it increases at high redshift.  In between, a peak
must obviously occur.  It is interesting to examine what determines
the location of this peak in our model.

\begin{figure}
\begin{center}
\resizebox{8.0cm}{!}{\includegraphics{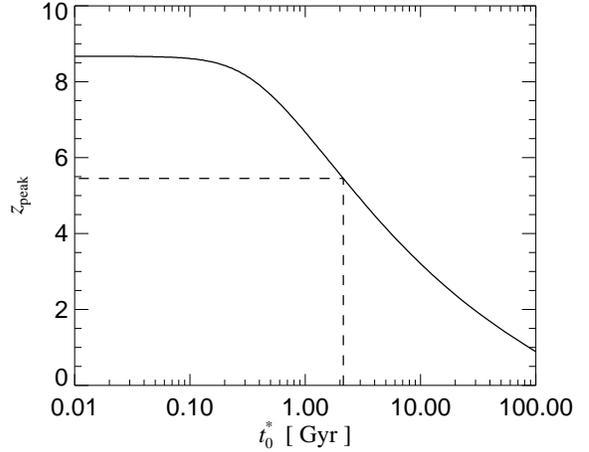}}%
\caption{\label{figpeak} The redshift position of the peak of the
cosmic star formation history as a function of the star formation
timescale $t^\star_0$ used in our multi-phase model. For our adopted
normalisation, which matches the observed star formation rates in
present-day spiral galaxies, the peak occurs at
redshift $z_{\rm peak}=5.45$. However, even if an arbitrarily short
gas consumption timescale is chosen, the peak cannot be pushed to
higher redshift than $z_{\rm peak}\simeq 8.7$.}
\end{center}
\end{figure}

At moderately high redshift, it is sufficient to consider
equation~(\ref{eqnbasicmodel}) as the prediction of our star formation
model, because then the extra contribution of star formation from
halos more massive than $10^{6.5}\,{\rm K}$ can be
neglected. Interestingly, the location of the maximum of this curve is
independent of $s_0$, but is somewhat sensitive to the prescribed
maximum of the normalised star formation rate in our model, as imposed
by the value of $\tilde\chi$.  If we assume that such a maximum does
not exist, i.e.~for $\tilde\chi\rightarrow\infty$, then the star
formation rate peaks at a redshift $z_{\rm peak}=8.67$, {\em
independent} on the details of our modeling of star formation. For any
finite value of $\tilde\chi$, the peak of the star formation history
will occur at a lower redshift.

This highlights that the exponential decline of the abundance of
star-forming halos at high redshift will always overwhelm any
power-law scaling of the star formation efficiency with expansion
rate, even if this scaling is very steep, as assumed here.
Consequently, it is not possible to push the peak of the star
formation history to arbitrarily high redshift.  In particular, if the
cooling rate is indeed limiting the star formation rate in the way
found here, the peak must occur below a redshift of $z\simeq8.7$ in
the $\Lambda$CDM cosmology.

We obtain further insight about the dependence of the peak's position
on model parameters by recalling that we were able to relate the
maximum normalised star formation rate to the gas consumption timescale
$t_0^\star$, where $t_0^\star$ is the free parameter of our hybrid
multi-phase model for star formation \citep{SprHerMultiPhase}. In
particular, we expect that $\tilde\chi$ will vary as
$\tilde\chi^{\frac{9}{2\eta}} \propto 1/t_0^\star$ when the model
parameter $t_0^\star$ is varied. For the simulations analysed here,
$t_0^\star$ was set to $t_0^\star=2.1\,{\rm Gyr}$ by normalising the
star formation rate of disk galaxies to the empirical Kennicutt Law
\citep{Ke89,Ke98} at $z=0$.

In Figure~\ref{figpeak}, we show the expected location of the peak of
the star formation history when $t_0^\star$ is modified with respect
to our fiducial choice of $2.1\,{\rm Gyr}$, with its corresponding
peak at $z_{\rm peak}=5.45$.  To delay the peak to a redshift as low
as $z=3$, the gas consumption timescale would have to be increased by
about a factor of 5 to an uncomfortably high value of $\simeq 10\,{\rm
Gyr}$. Note in particular that this would make us miss the
normalisation of the Kennicutt Law by about the same factor, and
would spoil our match of the observed density threshold for the onset
of star formation in disk galaxies.

\section{Derived quantities}  \label{secderived}

The cosmic star formation history directly determines a wide range of
key observables of the universe.  Making use of the computational
simplification provided by the analytic fit for the star formation
density derived above, we can conveniently obtain a number of such
predictions.  Clearly, this application is one of the main reasons why
an analytic closed-form description of the star formation history is
valuable. Among the range of direct implications of the star formation
history, we will here consider the stellar density amd metal
enrichment history of the universe, the observable supernova and GRB
rates on Earth, and the expected evolution of the density of compact
objects.

\subsection{Stellar density}

The mass density of long-lived stars that have formed at redshifts
higher than $z$ is simply given by \be \rho_\star(z) = \int_0^{t(z)}
\dot\rho_\star\,{\rm d}t = \int_{\chi(z)}^{\infty}
\dot\rho_\star(\chi) \frac{\chi^{1/2}}{H_0(\chi^3-\Omega_\Lambda)}
\,{\rm d}\chi. \ee Note that for the last equality we assumed a flat
cosmology.  If we express the stellar density in units of the baryon
density, we obtain the fraction \be f_*(z)=
\frac{\rho_\star(z)}{\rho_b}\ee of baryons locked up in stars at a
certain redshift, where $\rho_b= \Omega_b \rho_{\rm crit}$.  Using our
fitting function (\ref{eqnnewfit}) and the cosmological parameters 
of our simulations ($\Omega_b=0.04$, $\Omega_\Lambda=0.7$), we obtain
$f_*(0)=9.2\%$, in good agreement with our direct simulation result of
9.3\%, and consistent with observational constraints \citep{Cole2001,
Bal01,Fuk98} once the baryons in the warm-hot IGM \citep{Cen99,
Da99,Dave2001} are taken in account \citep{SprHerSFR}.
In Table~\ref{tabsfrcumul}, we give the redshifts and lookback
times for which the cumulative number of stars has reached a certain
fraction of the present day value.  These numbers are also in good
agreement with the corresponding simulation values given in
\citet{SprHerSFR}, confirming once more that the analytic fitting
function accurately describes our simulation results.

\begin{table}
\bc
\begin{tabular}{ccc}
\hline
Fraction & $z$ & $T\;[{\rm Gyr}]$ \\
\hline
  0.1 &  6.10 & 12.57  \\
  0.2 &  4.65 & 12.20  \\
  0.3 &  3.67 & 11.79  \\
  0.4 &  2.90 & 11.28  \\
  0.5 &  2.24 & 10.58  \\
  0.6 &  1.65 &  9.61  \\
  0.7 &  1.14 &  8.23  \\
  0.8 &  0.69 &  6.26  \\
  0.9 &  0.31 &  3.53  \\
 \hline
\end{tabular}
\caption{Cumulative star formation history as a function of lookback
time $T$ and redshift $z$. In each row, we list the times at which a
certain fraction $f_*(z)/f_*(0)$ of stars has formed.
\label{tabsfrcumul}}
\ec
\end{table}

\subsection{Metal enrichment \label{secmetenrich}}

Here, we make a rough estimate of the metallicity evolution of
the universe, assuming instantaneous recycling.  For
every mass element ${\rm d}M_\star$ of long-lived stars formed,
a gas mass equal to ${\rm d}M_{Z} = y \,{\rm d}M_\star$ is transformed
to heavy elements and returned to the interstellar or intergalactic
medium. Here $y$ is the yield, which we assume to be independent of
environment and epoch.

The metals deposited in the gas can either remain there, or they can
become permanently locked up in long-lived stars forming out of
enriched gas. If we define $\overline{Z}_\star$ as the mean
mass-weighted metallicity of all stars, and $\overline{Z}_{\rm gas}$
as the mean metallicity of {\em all} remaining gas, then all metals
produced up to a given epoch can be found either in stars or in the
gas. This metal budget can thus be written as $y\,\rho_\star =
\overline{Z}_\star \rho_\star +(\rho_b-\rho_\star) \overline{Z}_{\rm
gas}$, so that the mean mass-weighted metallicity of the gas follows
as \be \overline{Z}_{\rm gas}(z) =
\frac{\rho_\star(z)}{\rho_b-\rho_\star(z)} \left[ y -
\overline{Z}_\star (z)\right] = \frac{y - \overline{Z}_\star
(z)}{f_*^{-1}(z)-1}. \label{eqnenrichment} \ee If there is no loss of
metal-enriched gas from star-forming regions at all, we would expect
that the mean metallicity of the stars should be almost equal to the
yield $y$, the asymptotic value for a ``closed-box'' model. However,
in our simulations, star-forming galaxies are ``leaky'', and
particularly at low mass scales, they can loose metals efficiently by
galactic outflows. Observationally, there is also substantial evidence
that metals are able to escape from star-forming regions, as for
example shown by metal-absorption lines discovered in the low-density
intergalactic medium.  As a result, we expect $\overline{Z}_\star(z)$
to be -- perhaps considerably -- smaller than $y$. 

An accurate quantitative estimate of $\overline{Z}_\star(z)$ requires
a detailed modelling of the gas enrichment and transport processes
occurring during galaxy formation, which is beyond the scope of this
work.  However, we can make a simple estimate for the amount of metals
that can escape from small galaxies by winds. In the model
that we used in our simulation work, winds are generated with a fixed
velocity, and with a mass-loss rate equal to twice the star formation
rate. An escaping wind can hence be expected to transport about $2/3$
of the metals produced by the stars from the highly overdense
interstellar medium into the intergalactic medium (IGM). The 
remaining
1/3 will be locked up in long-lived stars. Whether or not a wind can
leave a star-forming galaxy depends to a large extent on the escape
velocity of its halo, which in itself depends directly on its virial
temperature. For simplicity, we here assume that winds can always
escape from halos with virial temperature below $10^{6.5}\,{\rm K}$,
while they stay completely confined in larger halos. These large
galaxies then act much like closed-boxes, with most of the metals
released by supernova ending up in long-lived stars.  In this simple
picture, we can then estimate the amount of metals in stars as \be
\overline{Z}_\star(z) \rho_\star(z) \simeq \frac{1}{3} y
\rho_\star'(z) + y[\rho_\star(z) - \rho_\star'(z)], \ee where
$\rho_\star'$ describes the density of stars that have formed in halos
of virial temperature below $10^{6.5}\,{\rm K}$. In this simple
estimate, we neglected the possibility of ``metal reaccretion'' from
the enriched IGM. Note that the rate $\dot\rho_\star'$ at which stars
form in halos smaller than $10^{6.5}\,{\rm K}$ can be obtained by
means of equation (\ref{eqnfullresult}) if we replace the number 2 in
the numerator of its last term by minus 1.  The estimated mean
metallicity of the gas then follows as $\overline{Z}_{\rm gas} = 2 y
\rho_\star / [ (f_*^{-1}-1) \rho_\star' ].$

\begin{figure}
\begin{center}
\resizebox{8.3cm}{!}{\includegraphics{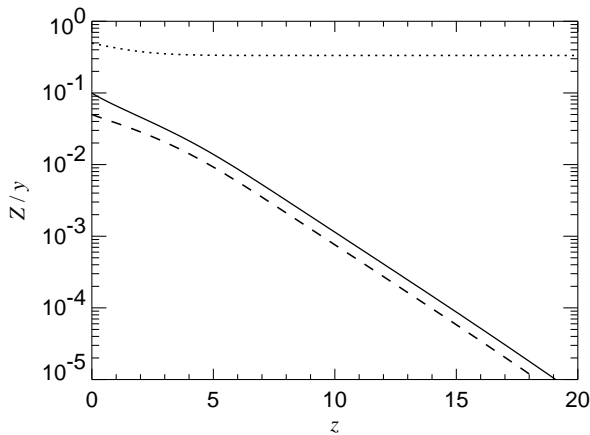}}%
\caption{\label{figenrichment} Upper limit for the mean mass-weighted
metallicity of ambient gas as function of redshift, normalised to the
assumed stellar yield $y$ (solid line).  The true mean metallicity will be lower by
the amount of metals locked up in stars forming from enriched
gas. Using a simple wind-escape model, we have also computed estimates for
the mean metallicities of stars (dotted) and gas (dashed) when the
leakage of metals from small galaxies by galactic winds is taken into
account.}
\end{center}
\end{figure}

Clearly, a detailed analysis of hydrodynamical simulations will be
required to check the accuracy of the above crude estimate.  However,
we note that neglecting $\overline{Z}_\star(z)$ in
equation~(\ref{eqnenrichment}) provides a strong upper limit for the
mean gas metallicity as a function of epoch.

In Figure~\ref{figenrichment}, we show the resulting upper limit
$\overline{Z}_{\rm gas}^{\rm max}(z) = y/ [f_*^{-1}(z)-1]$ as a
function of redshift, together with our estimates for the expected
mean metallicities of stars and gas in the framework of the above
wind-escape model.  For a solar yield of $y\simeq 0.02$, the mean
mass-weighted metallicity at $z=3$ could hence reach values of up to
$3.5\times 10^{-2}$ solar. However, since large galaxies are expected
to confine most of the metals they produce, locking them up in
long-lived stars, a more realistic estimate is the value of $\sim
2.0\times 10^{-2}$ solar derived for the wind-escape model.  Note,
however, that we expect strong spatial variations in the metallicity of
the gas, with a tendency of higher density regions to be more metal
rich.  The actually observed metallicity of gas of mean cosmic density
could hence be quite a bit lower than the above limits.

\subsection{Supernova and GRB rate}

For definiteness, we here assume a universal initial mass function
(IMF) of Salpeter (\citeyear{Sal55}) form with slope $-1.35$ in the
mass range $0.1\,{\rm M}_\odot$ to $40\,{\rm M}_\odot$. We further
assume that all massive stars above $8\,{\rm M}_\odot$ explode as
supernovae after a short lifetime of $T_{\rm sn}=3\times 10^7\,{\rm
yr}$. The number of supernovae per unit mass of long-lived stars is
thus given by \be f_{\rm sn}= \frac{\int_{8\,{\rm M}_\odot}^{40\,{\rm
M}_\odot} f(M) M^{-1} \, {\rm d}M} {\int_{0.1\,{\rm M}_\odot}^{8\,{\rm
M}_\odot} f(M)\, {\rm d}M} \simeq 7.9\times 10^{-3}\,{\rm
M}_\odot^{-1} ,\ee where $f(M)\propto M^{-1.35}$.  Note that the star
formation rate $\dot\rho_\star(z)$ we considered in our simulations
and in the analysis of this paper up to this point, is to be
understood as the rate at which long-lived stars form. This rate does
not include the formation rate of massive stars, which are instead
assumed to explode instantaneously and to return all of their mass to
the ambient gas.

The comoving number density of supernova explosions, $\dot n_{\rm
sn}$, is then simply expected to follow the star formation rate,
retarded by the lifetime of massive stars: \be \dot n_{\rm sn}(t)
=f_{\rm sn}\, \dot\rho_\star(t-T_{\rm sn}). \ee The retardation effect
can usually be neglected at low redshift, where the evolutionary
timescale of the star formation rate is large compared to $T_{\rm
sn}$.

We may also ask at which rate supernova events could be detected by an
observer on Earth, and what the redshift distribution of these events
would be. Defining the comoving distance to an event at an observed
redshift $z$ as \be d(z)=\int_0^z\frac{c\,{\rm d}z'}{H(z')}, \ee the
predicted rate of supernovae per unit redshift element and unit solid
angle is given by \be \frac{{\rm d}\dot N_{\rm sn}}{{\rm d}z\,{\rm
d}\Omega} (z)= f_{\rm sn} \dot\rho_\star(z') \frac{c\,\,
d^2(z)}{(1+z)H(z)}.  \ee Here $z'$ denotes the ``retarded redshift''
obtained by transforming $z$ to lookback time, adding the supernova
lifetime $T_{\rm sn}$ and transforming back to redshift. The factor
$(1+z)$ in the denominator takes care of the cosmological time
dilation effect.

\begin{figure}
\begin{center}
\resizebox{8.0cm}{!}{\includegraphics{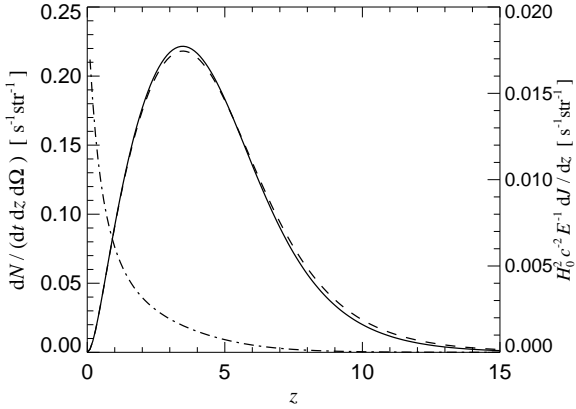}}%
\caption{\label{figsupernovarate} Distribution of the supernova rate
per unit redshift interval as -- in principle -- observable on Earth
(solid line).  The dashed line shows our prediction if the time delay
$T_{\rm sn}$ between formation and explosion of massive stars is
neglected. The dot-dashed line gives the redshift distribution of the
total specific intensity generated by the background of all
supernovae. }
\end{center}
\end{figure}

In Figure~\ref{figsupernovarate}, we show the redshift distribution of
this observable supernova rate. Interestingly, the cosmological
effects make the distribution peak at substantially lower redshift
than the star formation rate density itself. Over the full redshift range,
a total supernova rate of ${\rm d}\dot N_{\rm sn}/{\rm d}\Omega =
1.27\,{\rm s^{-1}str^{-1}}$ is predicted, or about 15.9 per second
over the whole sky. Unfortunately, most of these events will be too
distant and hence too faint to ever be observable.

\begin{figure}
\begin{center}
\resizebox{8.6cm}{!}{\includegraphics{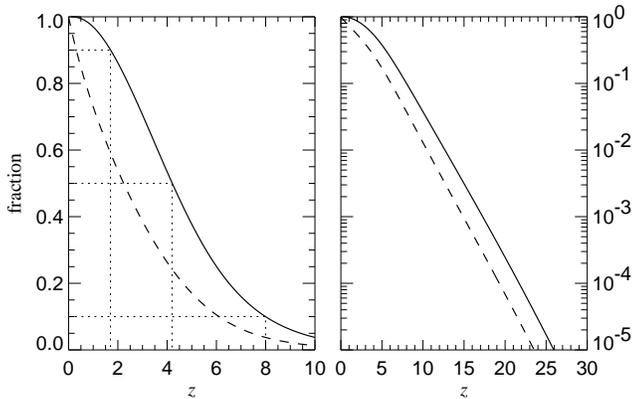}}%
\caption{\label{figgrbsupernovaintgrated} Fraction of all observable
GRBs/supernovae that occur at redshifts higher than $z$ (solid
lines). The left and right panel differ only in the scaling of the
y-axis. Also shown is the fraction $f_\star(z)/f_\star(0)$ of stars
that have formed prior to a given redshift.}
\end{center}
\end{figure}

Of course, the total energy flux received from supernovae is much more
peaked towards lower redshift than the event distribution itself. If,
for simplicity, a supernova is modelled as a standard candle with a
total bolometric emission of energy $E$ in radiation, then the
redshift distribution of the specific intensity of the total supernova
background radiation is given by \be \frac{{\rm d}J}{{\rm
d}z}(z)=\frac{c}{4\pi} E\frac{f_{\rm sn}\dot\rho_\star(z')}{(1+z)^2
H(z)}. \ee In Figure~\ref{figsupernovarate}, we have included a graph
that shows the redshift distribution of this flux.  About half the
total energy received on Earth from supernovae originated at redshifts
lower than $z=1.3$, but these events are only 5.5\% by number of all
supernova events that are in principle arriving on Earth.

There could also be a close relation between the supernova rate and
the rate of gamma ray bursts (GRBs) observed on Earth. While the
origin of GRBs is still one of the most interesting {\em open}
questions in cosmology, there are now a number of promising
theoretical models that link them to compact objects (black holes or
neutron stars) that form as end products of the evolution of very
massive stars. This suggests that the rate of GRBs should directly
follow the star formation rate, just like the supernova rate that we
considered above. Consequently, the observable GRB rate can be
computed in the same way as the supernova rate, with $f_{\rm sn}$
being replaced by the expected (but uncertain) number $f_{\rm grb}$ of
GRBs per unit-mass of long-lived stars.  \citet{Bromm02} have recently
given a detailed analysis of the expected GRB rate based on an
estimate along these lines.

In Figure~\ref{figgrbsupernovaintgrated}, we show the fraction of all
observable GRBs that originated at redshift higher than a given
epoch. Note that the corresponding fraction of supernovae is identical
if the time delay $T_{\rm sn}$ is neglected. For comparison, we also
give the integrated fraction $f_\star(z)/f_\star(0)$ of stars that
have formed up to a given redshift. Clearly, the distribution of
observable GRBs/supernovae is biased towards higher redshift compared
to the redshift distribution of all stars, even though the
GRB/supernova rate peaks actually at lower redshift than the star
formation history itself. About 50\% of the observable supernovae/GRBs
are expected to originate beyond redshift $z\simeq 4.2$, while it
takes until redshift $2.2$ before 50\% of all stars are formed.
\citet{Bromm02} have assumed a different star formation history with a
larger star formation rate at high redshift. Consequently, they
estimated a slightly higher redshift of about $z\simeq 5$ for the
epoch at which 50\% of the GRB signals have been generated.  Note
however that the GRB rate could be quite sensitive to variations of
the IMF with redshift, and to the possible existence of high-redshift
star formation mediated by molecular cooling, which we neglected here.

\subsection{Density in compact objects}

So far, we have assumed that stars below a mass of $8\,{\rm M}_\odot$
live essentially forever. However, especially the more massive ones
among them should have reached the end of their lifetime by the
present day, provided they have formed early enough in the history of
the universe.  Depending on their mass, they can then become
transformed into compact objects like white dwarfs or neutron stars,
for example.

We here want to compute a rough estimate for the fraction of stars
that have died since their formation time, but without going into the
complexities of full stellar evolution theory. We therefore
crudely assume that a star of mass $M$ has a lifetime of order \be
T_\star(M) \simeq T_\odot \left(\frac{M}{{ M}_\odot}\right)^{-2.3},
\ee where we put $T_\odot= 10\,{\rm Gyr}$ as the approximate
lifetime of a solar mass star. Neglecting any mass-loss processes
during the final stages of stellar evolution, the formation rate of
``compact objects'' is then given by \be \dot\rho_c(t) = \int_{0.1\,{\rm
M}_\odot}^{8\,{\rm M}_\odot} {\rm d}M\, f(M)\, \dot\rho_\star \left[
t- T_\star(M)\right] , \ee where $f(M)$ describes the normalised
Salpeter IMF. This quantity may also be viewed as an estimate of the
death rate of ordinary stars.

\begin{figure}
\begin{center}
\resizebox{8.0cm}{!}{\includegraphics{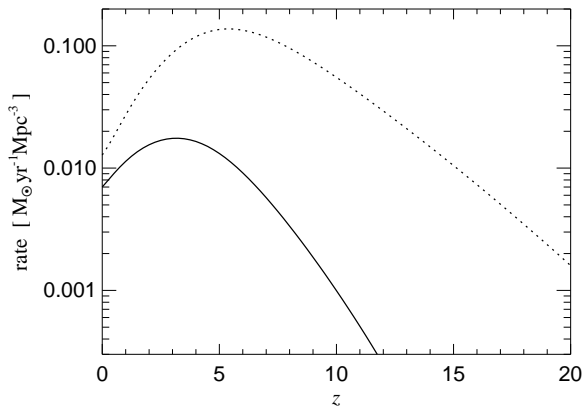}}%
\caption{\label{figcompactrate} Formation rate $\dot\rho_c$ of compact
objects as a function of redshift (solid line). For comparison, we
also show the formation rate of long-lived stars (dotted line).}
\end{center}
\end{figure}

\begin{figure}
\begin{center}
\resizebox{8.0cm}{!}{\includegraphics{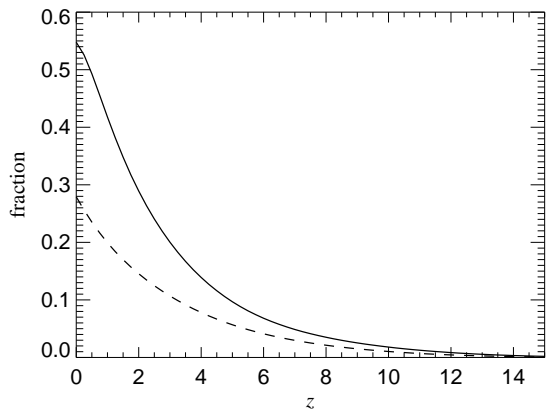}}%
\caption{\label{figcompactfraction} The ratio
$\dot\rho_c/\dot\rho_\star$ of the rate of stellar deaths to the star
formation density (solid).  At the present epoch, stars are already
dying with more than half the rate at which new stars are being
formed.  The dashed line gives the mass fraction of long-lived stars
that have turned into compact objects (i.e. that have reached the end
of their stellar lifetimes) as a function of redshift.}
\end{center}
\end{figure}

In Figure~\ref{figcompactrate}, we show the history of the
mass-weighted formation rate of compact objects.  As expected, the
death rate of stars peaks at lower redshift than the star formation
rate itself, which is simply a result of the dilation effect due to
the lifetime of stars. Note, however, that the mass-weighted rate of
stellar deaths is already strongly declining at present, which is a
non-trivial consequence of stellar lifetimes in relation to the cosmic
star formation history.  It is also interesting to compare the stellar
death rate to the star formation rate directly. In
Figure~\ref{figcompactfraction}, we show the ratio
$\dot\rho_c/\dot\rho_\star$ of the two as a function of
redshift. Interestingly, at the present epoch the rate of formation of
compact objects has reached about half of the rate at which new stars
are being formed, and $\dot\rho_c$ is set to exceed $\dot\rho_\star$
in the not too distant future.  In Figure~\ref{figcompactfraction}, we
also show the ratio of the cumulative density of stellar remnants to
the density of stars. At $z=0$, more than 25\% of all formed stellar
mass is expected to be in stars that have already reached the end of
their ordinary stellar lifetimes.

\section{Dependence on model parameters}   \label{seccosmology}

\subsection{Effects of cosmological parameters}  

The analysis of the physical origin of the cosmic star formation rate
carried out in Section~\ref{secphysbasis} allows us to
investigate expected variations in its evolution as a function of
cosmological parameters.  To this end, it is perhaps easiest to
consider equation~(\ref{eqnfullresult}), and to discuss the
cosmological dependence of the various terms involved.

The scaling of the background density $\overline{\rho}_0$ is simply
given by its definition; viz. $\overline{\rho}_0=3\Omega_0H_0^2/(8\pi
G)$.  Also, the dependence of $\chi(z)$ on cosmological parameters is
straightforward and follows from the usual scaling of the Hubble
constant.  For the normalised star formation rate $s_0\sim \left<\dot
M_\star\right>/M_{\rm vir}$, we expect it to scale like the cooling
rate normalised to the halo mass. Based on
equation~(\ref{eqncoolrate}), this implies \be s_0\propto (f_b
H_0)^{\frac{3}{\eta}}, \ee where $f_b=\Omega_b/\Omega_0$.  A slightly
more subtle question is how the parameter $\tilde\chi$, which limits
the star formation efficiency, depends on cosmological
parameters. Recall that the maximum  star formation rate we expect in a
halo of mass $M_{\rm vir}$ is given by $\simeq f_b M_{\rm
vir}/t_0^\star$ in the multi-phase model considered here.  On the
other hand, we assumed that this maximum is just attained as
$s_0\tilde \chi^{\frac{9}{2\eta}}$. If $t_0^\star$ depends only on
``local physics'' of the star forming gas, it should be approximately
independent of cosmological parameters. We thus expect $\tilde\chi$ to
scale as \be \tilde\chi = 4.6
\left(\frac{\Omega_b/\Omega_0}{0.133}\right)^{-0.3}
\left(\frac{h}{0.7}\right)^{-2/3}. \ee Finally, the shape and
normalisation of the power spectrum influence the evolution and
amplitude of $\sigma_4(z)$ and $\sigma_{6.5}(z)$, respectively.  In
particular, the high redshift behaviour of the star formation rate
will be quite sensitive to the amplitude of $\sigma_4(z)$. The lower
the amplitude, the larger the slope parameter $\beta$, implying a
faster decline of the star formation rate towards high redshift.

\begin{figure}
\begin{center}
\resizebox{8.0cm}{!}{\includegraphics{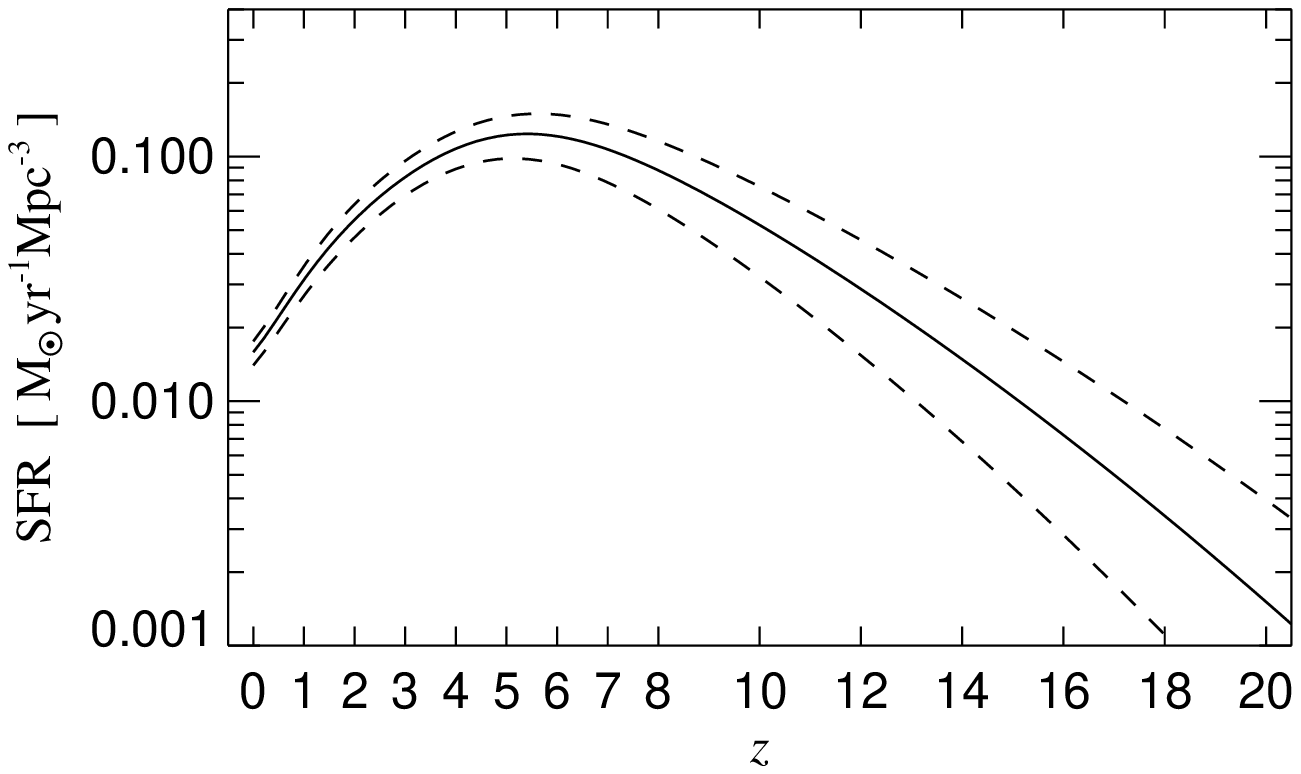}}\\%
\resizebox{8.0cm}{!}{\includegraphics{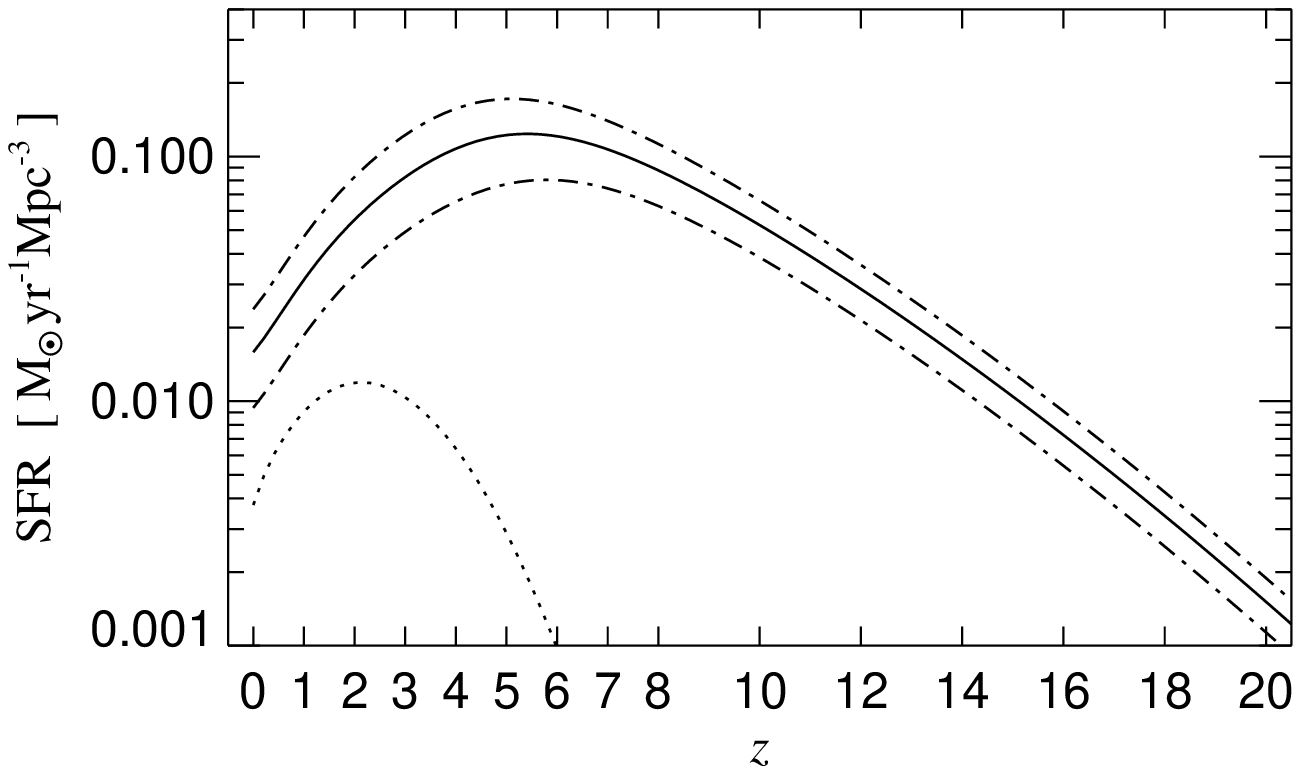}}\\%
\caption{\label{figcosmicdependence} Examples of the expected
dependence of the cosmic star formation history on cosmological
parameters. In the top panel, we show the star formation density for
the $\Lambda$CDM model when the normalisation of the power spectrum is
changed to $\sigma_8=1.0$ or $\sigma_8=0.8$, respectively (upper and
lower dashed lines). Similarly, the lower panel shows the effect when
the baryon density is varied to $\Omega_b=0.05$ or $\Omega_b=0.03$,
respectively (dot-dashed lines).  In both cases, the solid lines give
our standard result for the $\Lambda$CDM model, for
comparison. Finally, the dotted line shows the result for a $\tau$CDM
model of critical density, with baryon density $\Omega_b=0.08$, Hubble
constant $h=0.5$, and normalisation $\sigma_8=0.6$. }
\end{center}
\end{figure}

In Figure~\ref{figcosmicdependence}, we show a few examples of star
formation histories expected for different cosmological parameters.
For simplicity, we mainly restrict ourselves to simple variants of the
$\Lambda$CDM model. In particular, we show results for variations of
the power spectrum normalisation in the top panel, and for changes of
the baryonic density in the lower panel. It is clearly seen that the
high redshift star formation rate is particularly sensitive to the
normalisation of the power spectrum. This is also one of the reasons
why a $\tau$CDM model with critical density, which we show in the
lower panel of Figure~\ref{figcosmicdependence}, is expected to
feature a quite different star formation history with much less
high-redshift star formation, and a peak at substantially lower
redshift. The effects of the low normalisation of $\sigma_8=0.6$ of
this model are further amplified by the different evolution of the
growth factor as compared to the $\Lambda$CDM cosmology.

\begin{figure}
\begin{center}
\resizebox{8.0cm}{!}{\includegraphics{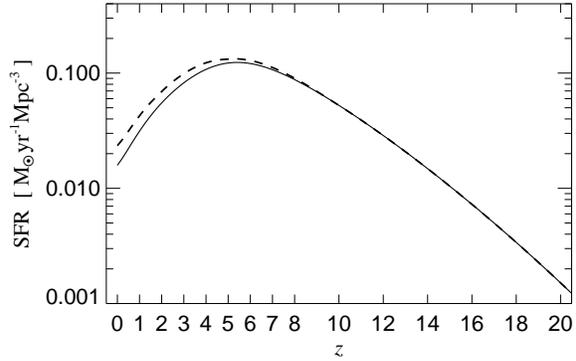}}\\%
\caption{\label{figmetaleffect} Estimate of the possible effect of
metal line cooling on the cosmic star formation history.  The dashed
line represents our prediction when the metallicity dependence of the
cooling function at $10^{6.5}\,{\rm K}$ is taken into account,
assuming a uniform metallicity following the predicted evolution of
the mass-weighted mean metallicity of the gas. The solid line
represents our default result of equation~(\ref{eqnfullresult}), for
comparison.}
\end{center}
\end{figure}

\subsection{Effects of metal line cooling \label{secmetalcooling}}

It is well known that the cooling rate of gas can be increased
substantially even by relatively little enrichment with heavy
elements. This is particularly true in the temperature range $\sim
10^{4.5}-10^{6.5}\,{\rm K}$, where an enrichment to solar metallicity
can increase the cooling rate by an order of magnitude, while even a
low metallicity of $10^{-2}\,{\rm Z}_\odot$ still enhances cooling by
roughly a factor of two. At higher temperatures, the sensitivity to metal
enrichment is significantly weaker, though.

So far, we have neglected the effect of metal enrichment on the
cooling efficiency of gas, both in our simulations and in our
present analysis.  This might result in a systematic underestimate of the
cooling and star formation rates, particularly at low redshift, where
the average metallicity of gas reaches potentially important
levels. However, whether the predictions we obtained in the framework
of our model are really altered by metal enrichment in a significant
way is less clear than it may seem, as we now discuss.

We first note that there are really two quite different regimes where
the cooling rate of gas is important. There is on one hand the diffuse
gas in galactic halos, which must dissipate its thermal energy
radiatively in order to collapse onto the highly overdense
interstellar medium (ISM) in the centres of galaxies. This is the
principal supply channel for gas that becomes newly available for star
formation.

On the other hand, there is the gaseous multiphase structure within
the ISM itself, where most of the baryonic mass is in cold clouds, but
a small fraction of it fills the volume as a hot intercloud medium,
heated by supernova explosions. This hot phase of the ISM is expected
to reach high metallicity quickly as soon as local star formation
starts, and it is hence in principal subject to very strong cooling
enhancement by metal lines. However, within the simple model that we
developed to describe the ISM, this metallicity effect can be offset
by a slight adjustment of the adopted temperature structure of the ISM
(as controlled by the evaporation efficiency $A_0$, see
\citeauthor{SprHerMultiPhase},~\citeyear{SprHerMultiPhase}), and to a
lesser extent, by a small change of the gas consumption timescale,
$t_0^\star$. The parameters can be chosen such that the normalisation
of the Kennicutt Law is maintained, yielding, to first order, an
unaltered dynamical behaviour of the ISM model. While it hence
may have been more consistent to assume high metallicity for the ISM
to begin with, this would not have changed our model predictions but
only the parameter values required to match the Kennicutt Law that we
used as a normalisation constraint.

We are thus primarily left with the influence of metal enrichment on
the cooling of gas within the diffuse atmospheres of halos.  At high
redshift, cooling is so efficient that we expect gas to cool nearly
instantly, even without any enrichment with metals.  In this regime,
the evolution of the star formation rate is driven by the fast
gravitational growth of the halo mass function, and we thus expect our
results to be largely independent of metal enrichment.

However, at low redshift, cooling becomes inefficient and the supply
of star forming gas is regulated by the cooling rate. Consequently, we
expect that the star formation rates would be higher if the cooling
gas is significantly enriched with heavy elements. Unfortunately, it
is far from clear how efficiently metals that are expelled from the
star forming ISM of galaxies are mixed with the rest of the gas in the
universe.  While small galaxies can efficiently expel metals with
galactic winds into the IGM, where they are stopped by shocks, the
resulting bubbles of metal-enriched gas may be too hot to be accreted
again by similarly small galaxies. Gas that cools onto these small
galaxies may then always end up being mostly pristine. On the other
hand, the metals deposited in the IGM may be ``recollected'' in the
collapse of significantly larger objects, for example in galaxy groups
or clusters, resulting in pre-enriched halos. In general, we expect
larger halos to have a better chance of accreting metal-enriched
gas. Note, however, that the relative enhancement of the cooling rate by
metals becomes weaker for larger halos due to their higher virial
temperatures.

Perhaps the simplest model we can make for estimating the possible
influence of metal-line cooling on our prediction for cosmic star
formation is to assume that the diffuse gas cools with the average gas
metallicity that we estimated in section~\ref{secmetenrich}. For
simplicity, we neglect the temperature dependence of the cooling
enhancement at a given metallicity, and instead approximate it with
the values appropriate for halos of virial temperature $\sim
10^{6.5}\,{\rm K}$. Halos with this temperature or higher dominate the
star formation rate at low redshift. For halos larger than
$10^{6.5}\,{\rm K}$, we will then overestimate the cooling enhancement
effect, but this is offset to some extent by a possible underestimate
for smaller halos, where the dependence of the cooling rate on
metallicity is stronger.

In Figure~\ref{figmetaleffect}, we show the resulting estimate for the
evolution of the star formation rate when the enrichment history is
taken into account by a global increase of the value of the cooling
function for all halos, assuming a yield of $y=0.02$. Note that we
found earlier that the cooling rate scales slightly weaker than linear
with the cooling function, as $\propto [\Lambda(T,
Z)]^{(3-\eta)/\eta}$, which alleviates the metallicity dependence
slightly. Nevertheless, we obtain an estimated increase of the star
formation density of about $50\%$ at $z=0$. Because this metallicity
effect becomes weaker towards higher redshift, the increase in the
total stellar density is only about 25\%.

Hence, metals have the potential to alter the star formation
history at low redshift. However, more reliable
estimates of the strength of this effect require a better
understanding of the mixing processes of ejected metals with the
gas. Depending on the details of these processes, the effects of metal
enrichment may be more or less substantial than estimated here. Note that
the importance of metal enrichment effects is also intimately linked
to the strength of galactic winds, or more generally, to the physical
nature of feedback processes.  We remark that without the inclusion of
winds, accreted gas from the IGM would always be pristine in our
simulations.

\section{Discussion}  \label{secconclusions}

We have formulated an analytical model to identify physical processes
that play an important role in determining the evolution of the cosmic
star formation rate density, $\dot\rho_\star (z)$.  Using this model, we
obtain simple closed-form expressions for $\dot\rho_\star (z)$ which
match hydrodynamic simulations that include star formation and
feedback to a level of $\approx 10 \%$.  Our model, therefore,
provides a framework for interpreting both theoretical and
observational estimates of $\dot\rho_\star (z)$.

Our analysis shows that the evolution of the cosmic star formation
rate is characterised by a number of generic features in hierarchical
universes.  These properties depend on cosmological parameters but are
largely insensitive to the detailed physics of star formation.

In particular, we have identified two broad regimes of star formation
that are separated by a peak in $\dot\rho_\star (z)$ at $z=z_{\rm peak}$.
At high redshifts, $z > z_{\rm peak}$, cooling is very efficient and
halos contain abundant quantities of star-forming gas.  In this regime,
the dominant contribution to the global star formation rate comes
from the highest mass halos present at any time that are not 
unusually rare.  Consequently, $\dot\rho_\star (z)$ follows the
evolution of the exponential part of the halo mass function.  The
logarithmic slope of this phase of evolution depends on properties
of the cosmology but not on the details of star formation, which
only affect the overall normalisation.

At low redshifts, $z < z_{\rm peak}$, cooling becomes inefficient, and
the supply of star-forming gas is regulated by the cooling rate.  In
this regime, $\dot\rho_\star (z)$ gradually declines from its maximum
at $z = z_{\rm peak}$ to $z=0$ as a power-law function of the
expansion rate, $\dot\rho_\star (z) \propto H(z)^q$.  Typically, we
find $q\approx 4/3$, weakly dependent on the gas density profiles
within dark matter halos. The scaling may also be altered slightly if
metal enrichment becomes important in halos at late times.  To the
extent that our results apply to the real Universe, observations of
$\dot\rho_\star (z)$ at $z < z_{\rm peak}$ should be well-fitted by a
functional form $\dot\rho_\star (z) \propto H(z)^q$.  Thus, our
prediction for the evolution of the cosmic star formation rate is, in
principle, testable by accurate measurements of $\dot\rho_\star (z)$
at low redshifts.

We have shown that the existence of a peak in the star formation rate
at a redshift $z = z_{\rm peak}$ is generic, but that the value of
$z_{\rm peak}$ depends on assumptions about the characteristic gas
consumption timescale, as parameterised by $t_0^\star$.  For plausible
values of $t_0^\star$ we find that $z_{\rm peak}$ should be restricted
to the range $3 \simlt z_{\rm peak} < 8.7$, with a firm upper limit
corresponding to instantaneous gas consumption.  In our numerical
simulations, in which we chose $t_0^\star$ to reproduce the empirical
Kennicutt Law, $z_{\rm peak} \approx 5.5$.

Overall, we broadly predict that the cosmic star formation history in
hierarchical universes should have a generic form, rising
exponentially at first, peaking at $z_{\rm peak}$, and then declining
to $z=0$ as a power-law function of $H(z)$.  The logarithmic slopes on
either side of the peak are mainly determined by cosmology, but the
overall normalisation of $\dot\rho_\star (z)$ and the value of $z_{\rm
peak}$ are sensitive to assumptions about gas and star formation
physics.  The generic form we propose is compactly summarised by
e.g. equation~(\ref{eqnnewfit}), where the value of $\beta$ is fixed
by cosmology and $\dot\rho_\star(0)$ and $\alpha$ determine the
overall normalisation and the location of $z_{\rm peak}$.

We note that there are implicit assumptions in our description of the
star formation physics that can influence e.g. $\alpha$ and $z_{\rm
peak}$ in fits of the form of equation~(\ref{eqnnewfit}).  For
example, for simplicity we have assumed that cooling rates are those
appropriate for a H/He plasma of primordial abundance.  It is believed
that at very high redshift, $z\sim 20 - 30$, molecular cooling is an
important physical mechanism for early star formation (e.g. Bromm et
al. 1999, Abel et al. 2002).  Thus, our results are not applicable to
Population III (e.g. Carr 1984) star formation and will not properly
characterise $\dot\rho_\star (z)$ until star formation globally
resembles that at lower redshifts.  How and when the Universe made
this transition is uncertain.  Simple estimates suggest that it may
have occurred at $z\sim 15-20$, as the Universe began to become
chemically enriched (e.g. Mackey et al. 2002), but more detailed
studies of this fundamental issue are clearly needed.

In our simulations, we have also ignored the contribution of metal
line cooling to the overall cooling rate.  However, we have been able
to estimate the possible importance of this process using our
analytical model, as described in Section~\ref{secmetalcooling}.
While our conclusions depend in detail on uncertainties in how
efficiently enriched gas is mixed into galactic halos and the IGM, we
find that metal line cooling does not affect our results at early
times and likely has only a modest influence on the behaviour of
$\dot\rho_\star (z)$ at low redshift.  In particular, if metals
carried by galactic winds are efficiently mixed with the remaining gas
in the universe, enhancements in the cooling and star formation rates
at low redshift increase the stellar density only by $\sim 20 - 30\%$
by $z=0$, boosting the star formation rate by a similar factor for
$z\simlt 3$.  We note that when we previously compared our simulations
to observations, there was an indication that our predicted
$\dot\rho_\star (z)$ was perhaps low at $z\sim 1$ (e.g. figure 12
in Springel \& Hernquist 2002b).  If we take this discrepancy
seriously, given observational uncertainty, then metal cooling 
would boost the predicted star formation rate into the observed range,
at least according to our present simple estimates, without violating
constraints on the total stellar density at $z=0$.

Metal enrichment can also, in principle, influence the location of
$z_{\rm peak}$ and shift it to lower $z$.  However, our current
estimate of this effect suggests that the peak will still lie at
a relatively high redshift, $z_{\rm peak}\sim 5$, as
indicated by Figure~\ref{figmetaleffect}, and would thus not
substantially alter our predictions for the evolution of the
stellar density or the mean age of the stellar population
with redshift.  Furthermore, we do not believe that the overall
form of the star formation history we considered here would be altered
significantly.  However, detailed hydrodynamical simulations of metal
enrichment processes will ultimately be required to more accurately
constrain the relevance of metal cooling for our modeling.

\section*{Acknowledgements}

We thank Simon White for instructive discussions and critical comments
that were helpful for the work on this paper.  This work was supported
in part by NSF grants ACI 96-19019, AST 98-02568, AST 99-00877, and
AST 00-71019.  The simulations were performed at the Center for
Parallel Astrophysical Computing at the Harvard-Smithsonian Center for
Astrophysics.

\bibliographystyle{mnras}
\bibliography{shape}

\end{document}